# New model of chromite and magnesiochromite solubility in silicate melts

Nail R. Zagrtdenov[a,b*], Michael J. Toplis[c], Anastassia Y. Borisova[b,d],

Jérémy Guignard[c]

[a] *CanmetMINING, Natural Resources Canada/Ressources naturelles Canada*

[b] *Géosciences Environnement Toulouse, Université de Toulouse, UPS, CNRS, IRD, CNES, Toulouse, France*

[c] *Institut de Recherche en Astrophysique et Planétologie, Université de Toulouse, UPS, CNRS, CNES, Toulouse, France*

[d] *Geological Department, Lomonosov Moscow State University, Vorobievu Gory, 119899, Moscow, Russia*

[*]Corresponding author: E-mail: **nailzagrtdenov@gmail.com**;

[*]Corresponding address: CanmetMINING, Natural Resources Canada/Ressources naturelles Canada, 555 Booth Street, Ottawa, Ontario, Canada K1A 0G1

# Abstract

Chromite is a key magmatic mineral frequently used as petrogenetic indicator of physico-chemical conditions of mafic magma crystallization. In this work, magnesiochromite and chromite solubility in a natural basalt and an iron-free haplobasalt was investigated at 1440°C and atmospheric pressure under controlled CO-$CO_2$ gas mixtures corresponding to the range two log units below to two log units above the fayalite-magnetite-quartz buffer (FMQ-2 to FMQ+2). The source of chromium was either natural chromite or synthetic $Cr_2O_3$, the latter reacting with the basaltic liquids to form a magnesiochromite. The highest concentrations of Cr in haplobasaltic melts are found in equilibrium with magnesiochromite, ranging from 3500 to 6800 ppm depending on redox conditions. In detail, at low $f_{O2}$, liquids have high Cr contents, but the variation of log [Cr, ppm] is not a linear function of log $f_{O2}$. Using our new data and data from the literature a model for Cr concentrations at chromite/magnesiochromite saturation in silicate melts has been developed based upon a thermodynamic formalism. This model has the form:

$$\frac{X^{liq}_{Cr^{tot}}}{X^{Chr}_{Cr_2O_3}} = exp(a + b*\lambda + c/T) * (1 + f_{o2}^{-\frac{1}{4}} * exp(d + k/T + g*\lambda)),$$

where $X^{liq}_{Cr^{tot}}$ is mole fraction of chromium in the liquid, $X^{Chr}_{Cr_2O_3}$ is mole fraction of chromium oxide in the chromite, T is temperature in Kelvins, $f_{o2}$ is oxygen fugacity in bars, $\lambda$ is optical basicity, and a (-7.01 ± 2.10), b (13.72 ± 2.73), c (-12405 ± 1253), d (24.46 ± 3.13), k (-24395 ± 2037), g (-23.59 ± 4.20) are constants. This model may be used to assess the effect of melt composition on chromite and magnesiochromite solubility in silicate melts during peridotite melting and assimilation. At moderate and high oxygen fugacities (for example, >FMQ-1 at 1200 °C), concentration levels of Cr at chromite saturation are higher for ultramafic than for felsic rocks. Our data imply that assimilation of magnesiochromite-bearing serpentinite lithosphere could result in high Cr contents in mafic melts, triggering massive crystallization of chromite, especially upon the system hydridization, cooling, oxidation and magma degassing. Our model may be applied for quantitative prediction of chromite crystallization and formation of the terrestrial mantle-crust transition zones and mantle represented by chromitites and dunites.

# 1. Introduction

Crystallization of chromite from fluids and silicate melts is the most probable origin of large concentrations of chromite in igneous rocks such as disseminated, stratiform, nodular and podiform chromitites (Roeder & Reynolds, 1991; Borisova et al., 2012; Johan et al., 2017). Experimentally determined chromite solubilities in silicate melts are variable (Maurel and Maurel, 1982b; Murck & Campbell, 1986; Roeder & Reynolds, 1991; Forsythe & Fisk, 1994; Poustovetov & Roeder, 2000) ranging from 50 ppm at 1150 °C and $f_{O_2}$ ~ -6.9 (Forsythe & Fisk, 1994) to 9800 ppm at 1300 °C and $f_{O_2}$ ~ 12.8 (Roeder & Reynolds, 1991), suggesting an importance of such physical-chemical parameters as temperature, oxygen fugacity and the melt composition controlling the chromite saturation in the silicate melts. High concentration of chromium at chromite saturation (~1220 ppm Cr) of basaltic melt at FMQ and 1350°C contrasts with the fact that natural basaltic to boninitic melts have lower concentrations of Cr (**Fig. 1**), with typical values in the range from 20 to 700 ppm (Gale et al., 2013); yet, the real Cr concentrations may be affected by the secondary fluorescence effect in the glasses (Borisova et al., 2018), although the effect is absent in the glasses without any traces of the Cr-rich minerals. The fact of wide variation of Cr in the natural mafic melts implies that natural melts of mantle origin have lost their initial Cr concentrations upon chromite crystallization during partial melting of peridotites and the melt percolation in the lithosphere and asthenosphere, magma mixing and possible hydridization (e.g., Borisova et al., 2020a,b).

The available experimental data suggest that the Cr content of melts saturated in chromite strongly depends on temperature and oxygen fugacity and, to a lesser extent, melt composition. For example, chromium concentrations in mafic melts increase with increasing temperature and rise dramatically when oxygen fugacity decreases (Maurel and Maurel, 1982b; Murck & Campbell, 1986; Roeder & Reynolds, 1991). The high Cr concentrations in silicate melts have long been postulated to be the result of assumed $Cr^{2+}$ which prevails in basaltic melts compared to $Cr^{3+}$ (Roeder & Reynolds, 1991), as subsequently confirmed by Berry & O'Neill (2004) and Berry et al. (2006) by X-ray Absorption Near Edge Structure spectroscopy (XANES). Concerning melt composition, basalts with higher MgO contents and lower $Fe_2O_3$ and $Al_2O_3$ contents have higher Cr contents at the chromite saturation, while more total alkalis and more FeO lead to lower Cr content (Murck & Campbell, 1986; Roeder & Reynolds, 1991). The latter observation was taken to indicate that $Cr^{2+}$ and $Cr^{3+}$ occupy similar structural positions in the melt as $Fe^{2+}$ and $Fe^{3+}$ leading to competition between these cations. The composition of chromites coexisting with silicate melts is also controlled by physico-chemical conditions, but the compositional $Cr^{3+}$ to $Al^{3+}$ ratio of the melt has the principal control (e.g. Maurel and Maurel, 1982b; Roeder & Reynolds, 1991). For this reason, at higher temperature, chromite has higher content of chromium oxide and MgO (Maurel and Maurel, 1982b; Murck & Campbell, 1986; Roeder & Reynolds, 1991). From a thermodynamic standpoint, an alternative way to look at chromite saturation is to consider the distribution coefficient of Cr between spinel and liquid ($D_{Cr}$). Maurel & Maurel (1982b) showed that this parameter increases when temperature decreases and oxygen fugacity increases. Roeder & Reynolds (1991) have showed that the distribution coefficient ($\frac{(Cr+Al+Fe^{3+})\text{ in the chromite}}{(Cr+Al+Fe^{3+})\text{ in the coexisting melt}}$) reaches a maximum of 140 at log $f_{O2}$= -5.0. The authors suggested that this point probably corresponds to the maximum of $Cr^{3+}$ content in the melt, more oxidized conditions leading to the formation of $Cr^{6+}$ in the melt (Schreiber, 1976). In other words, a mixture of $Cr^{2+}$ and $Cr^{3+}$ is inferred to dominate at log $f_{O2}$ lower than -5.0 (Roeder & Reynolds, 1991), consistent with direct measurement of chromium speciation ($Cr^{2+}/\sum Cr$) in synthetic silicate glasses as a function of oxygen fugacity (Berry & O'Neill, 2004). Indeed, at 1400 °C in the interval of oxygen fugacities characteristic for most mafic magmas (between nickel-nickel oxide and quartz-fayalite-magnetite buffers), $Cr^{2+}/\sum Cr$ ratio ranges from 0.3 to 0.8 as a function of melt composition, indicating a significant proportion of $Cr^{2+}$ in the melt (Berry & O'Neill, 2004). In terms of compositional dependence, Berry et al. (2006) found a linear correlation between log[$Cr^{2+}/Cr^{3+}$] and theoretical *optical basicity* (a compositional parameter, calculated from composition of the melt) (Duffy, 1993). Optical basicity is useful for ranking a wide variety of glasses in order of increasing basicity (high optical basicity in more mafic liquids). It correlates such properties as refractive index or redox equilibria irrespective of the anionic composition of the glass (Duffy, 1989). Higher optical basicity corresponds to lower $Cr^{2+}/\sum Cr$ ratio and iron-bearing and iron-free melts appear to follow the same trend.

To extend this previous work to wider range of silicate melts, which are saturated in chromite (($Fe,Mg,Mn)^{2+}(Cr,Al)^{3+}_2O_4$) or magnesiochromite ($MgCr_2O_4$), we have performed additional experiments of chromite saturation in anhydrous iron-containing and iron-free melts as a function of oxygen fugacity, also covering a range of Cr/Al ratios. These new data and those of the literature are used to develop a predictive thermodynamically based model of chromite saturation that may be applied to model the geochemistry of chromium in natural melts generated and evolved in the mantle and the mantle-crust transition zone.

## 2. Experimental and Analytical Methods

### 2.1. Starting materials

The starting material compositions are shown in **Table 1**. Two crystalline starting materials were used for this study. The first was natural chromitite from the Silesia ophiolite in Poland (S2 chromite, Wojtulek et al, 2016). This sample was crushed; and a fraction of 300 - 500 µm-size was separated by sieving. The chromite fraction was then separated from secondary silicates (e.g., serpentine) manually. After extraction of the chromite fraction, the chromite grains were leached using 6.2 N HCl and 5% HF, following the method of Snow et al. (1994). First, chromite was maintained in the leaching solution for 10 minutes, then the solution was removed. Second, chromite was kept in a new leaching solution in an ultrasonic bath for 10 minutes. Third, it was placed in a drying oven at 125 °C for 10 minutes in a closed Savillex container. After cooling, the solution was removed. After leaching, the chromite grains were rinsed with deionized water and dried. Minor residual salts were visible on the chromite surface, but their influence is considered negligible on the experimental results at high temperature. The second crystalline starting material was crystals of synthetic $Cr_2O_3$, extracted from an electromelted furnace refractory with a measured starting composition of 98.5 wt% $Cr_2O_3$, and 1.5 wt% $TiO_2$. This material was crushed; and the 300 - 500 µm-fraction was separated.

These crystalline materials were mixed with either a natural mid-ocean ridge basaltic glass (from Mid-Atlantic Ridge, Knipovich ridge, № 3786/3 in the 38$^{th}$ cruise of scientific research vessel Akademik Mstislav Keldysh; Sushchevskaya et al., 2000) that had been crushed to powder in an agate mortar to produce particles <100 µm in size or a synthetic iron-free haplobasaltic glass corresponding to the eutectic composition of the anorthite-diopside system ($Di_{63}An_{37}$). This material was made from reagent grade oxides ($SiO_2$, $Al_2O_3$) and carbonates ($CaCO_3$ and $MgCO_3$) that were melted at 1500°C, quenched and then crushed. This procedure was repeated twice to ensure homogeneity of the final glass powder.

## 2.2. Experimental method

Experiments were performed in a vertical gas-mixing furnace at one atmosphere pressure, 1440°C and controlled oxygen fugacity (Toplis et al., 1994). Redox conditions in the furnace were controlled by mixtures of CO and $CO_2$ spanning the range from two log units below to two log units above the fayalite-magnetite-quartz (FMQ) buffer (FMQ-2 to FMQ+2). The samples were held on Pt wire loops (the wire 0.3 mm in diameter and the loop typically 1 to 2 mm in diameter). For experiments using MORB at or below the FMQ buffer, the wire loops were presaturated with iron at the relevant $f_{O2}$ for 24 hours at 1400°C using an Fe-rich basalt powder (Toplis et al., 1994). These wire loops were cleaned using hydrofluoric acid and rinsed with MQ water to get rid of residual salts.

Four different mixtures were prepared for the equilibrium experiments. For the first two, 5 wt% of chromite crystals were mixed with the 95 wt% of mid-ocean ridge basalt MORB powder or $Di_{63}An_{37}$ powder. For the second two, 3 wt% of $Cr_2O_3$ crystals were mixed with 97 wt% of MORB or $Di_{63}An_{37}$ powder.

Droplets of these mixtures were attached to the wire loops during a 30 s heat treatment at 1400°C in air in a muffle furnace. Once samples of the four different starting mixtures were ready they were mounted together on a Pt-basket and then introduced into the hot-spot of the gas mixing-furnace, the basket being suspended at the end of a ceramic rod on a thin (0.2 mm Pt wire. The experimental run duration was estimated according to a minimal distance necessary for Cr to diffuse through the whole volume of thee reacting basaltic melt (e.g., Zagrtdenov et al., 2015). At the end of the experiments with duration of 24h at 1440°C, an electric current was passed through the thin wire causing it to melt and the basket to fall into the cold zone of the furnace, still under CO-$CO_2$ mixture. Quench rate with this method is estimated at 1000°C/s. After evacuation of the CO, the samples were recovered and mounted in epoxy and then polished by SiC grinding paper. The attainment of equilibrium was controlled by: i) homogeneity of the real Cr concentrations obtained by EPMA and LA-HR-ICP-MS and ii) the euhedral morphology of the magnesiochromite crystals in contact with the glass.

## 2.3. Electron Probe Microanalysis (EPMA)

Major elements in the crystals and glasses (and trace Cr in glasses) were analyzed using a CAMECA SX-Five electron microprobe at the Centre de Microcaractérisation Raimond Castaing (Toulouse, France). The glass analyses were performed at the distance more than 200 microns from the chromite grains in order to avoid the secondary fluorescence effect of the high-concentrated chromium-bearing phase (Borisova et al., 2018). An accelerating voltage of 15 kV and current of 20 nA were used. Synthetic and natural standards were used for calibration: albite (for Na), corundum (Al), wollastonite (Si, Ca), sanidine (K), pyrophanite (Mn, Ti), hematite (Fe), periclase (Mg), $Cr_2O_3$ (Cr). Element and background counting times for most elements were 10 s, whereas the counting time for Na and K was 5 s to avoid volatilization under the electron beam. The peak counting time for Cr was 120 s with a resulting detection limit of 70 ppm. MPI-DING mafic and ultramafic glasses (KL2-G, ML3B-G, GOR132-G, GOR128-G, Jochum et al., 2006) were analyzed as unknown samples to monitor the accuracy of the major and trace element analyses. The accuracy estimated on the reference glasses ranges from 0.5 to 3 % (1σ RSD = relative standard deviation), depending on the element contents in the glasses. Error bars (one standard deviation, characterizing homogeneity of the sample) are below 5 RSD% (relative standard deviation).

## 2.4. Method of laser ablation inductively coupled plasma mass spectrometry (LA-HR-ICP-MS)

To confirm no influence of the secondary fluorescence effect on the obtained Cr contents in the glasses analyzed by EPMA, additional major and trace (Cr, Co and Ni) element analyses were performed using a New Wave Research UP213 system (Fremont, CA, USA). This is a nanosecond laser system equipped with a Nd:YAG, Q-switched laser operating at 213 nm and a pulse width of 4 ns coupled with a high-resolution ICP-MS Element-XR. A 10 µm spot size was applied with 10 Hz and 30 $J/cm^2$ conditions. The background was measured during 30 s, ablation for 30 s and wash-out after ablation for 15 s. To quantify the elemental composition of the glasses, average calcium concentrations (based on $^{43}$Ca) measured by electron microprobe were used as an internal standard. SILLS 1.2.0. software was used for the elemental concentration quantification. For external calibration in bracketing mode, we used NIST SRM 610 reference material (Jochum et al., 2011). Furthermore, to control the accuracy, we used the MPI-DING mafic and ultramafic glasses (KL2-G, ML3B-G, GOR132-G, GOR128-G) (Jochum et al., 2011; Borisova et al., 2012) as secondary standards. The detection limit for major elements varied between 1 and 20 ppm, whereas it was 0.21 ppm for $^{53}$Cr, 0.04 ppm for $^{59}$Co, and 1.65 ppm for $^{60}$Ni. The relative accuracy for Cr estimated from the reference glasses ranges from 3.1 to 8.6 % (1σ RSD = relative standard deviation), depending on the element contents in the reference glasses.

## 3. Results

The data on composition of chromite crystals after experiments and chromium distribution coefficients between the chromite crystals and glass are shown in **Table 2**. The glass compositions after experiments are provided in **Table 3**. The description of the four systems is given below.

### 3.1. Chromite-MORB melt system

After dissolution of chromite in MORB, we find that after heat treatment, residual chromite crystals form a spongy texture (**Fig. 2**). The formation of this texture indicates that the basaltic melt partly dissolved the chromite crystals. Of note is the fact that during interaction with the melt, chromite changes in composition, becoming more Cr-rich. In detail, chromium oxide ($Cr_2O_3$) contents in the final chromite range from 49.67 wt% at FMQ+2 to 55.42 wt% at FMQ-2, compared to the initial average of 41 wt% $Cr_2O_3$. Total iron oxide content ($FeO_{tot}$) in the chromite is 12 wt% at FMQ-2. At higher oxygen fugacities (at FMQ and FMQ+2), it is significantly higher (16 and 17 wt%, respectively). Magnesium content (MgO) is 14.5 wt% at FMQ-2, 13 wt% at FMQ and 15 wt% at FMQ+2. Titanium oxide concentration is constant around 0.4 wt% for all redox conditions. Aluminium oxide contents range from 14.7 to 16.4 wt% $Al_2O_3$ that is dramatically lower than that of initial chromite (26 wt% $Al_2O_3$).

Alumina concentration is 16 ~wt% $Al_2O_3$ at all redox conditions, slightly higher than the initial 15.3 wt%, consistent with the reequilibration and dissolution of Al-rich chromite. The concentration of $TiO_2$ is close to 1.5 wt% (which is slightly higher than the starting 1.45 wt% content) with no dependence on oxygen fugacity. MgO content is around 9 wt% with no significant variation as a function of oxygen fugacity. Iron oxide content shows an irregular variation as a function of $f_{O2}$ (6.04 wt% at FMQ-2, 8.9 wt% at FMQ, and 7.2 wt% at FMQ+2), these complex variations possibly being the result of competing effects of chromite dissolution/reequilibration.

As illustrated in **Figure 3** the highest concentration of Cr (6300 ppm) is observed at the most reducing conditions (FMQ-2, $logf_{O2}$= -7.9) decreasing to 2856 ppm at the FMQ buffer and 1987 ppm at FMQ+2. This variation is not a linear function of log $f_{O2}$. **Figure 4** shows how the distribution coefficient of chromium (ratio of Cr concentration in the crystal to that in the liquid) varies as a function of oxygen fugacity.

## 3.2. Chromite-haplobasaltic melt system

Chromites dissolved in the haplobasaltic glass ($Di_{63}An_{37}$) also have a spongy texture similar to that described in the MORB glass. The chromium oxide content in final chromite is 63.8 wt% at FMQ-2 and 65.4 wt% at FMQ+2. These concentrations are slightly higher than those of chromite observed in the MORB melt, and much higher than for those of the initial chromite (41 wt%). Total iron oxide content ($FeO_{tot}$) in the residual chromite dissolved in haplobasalt ($Di_{63}An_{37}$) is 1.73 wt% at FMQ-2 and 1.06 wt% at FMQ+2 that is dramatically lower than the initial chromite. Magnesia concentration at both oxygen fugacities is around 21 wt% MgO, which is more than in the initial chromite (15 wt% MgO). Alumina content is 11.5 wt% $Al_2O_3$ at FMQ-2 and 10.3 wt% at FMQ+2. These values are much lower than the initial value of 26 wt% $Al_2O_3$.

Magnesium content is constant at all $f_{O2}$, with a value of ~15 wt% MgO (compared to the starting value of 11.3 wt%). Total iron oxide content varies from 0.22 wt% to 1 wt% in the final glass (that was initially iron-free). Chromium content in the haplobasaltic glass is depicted in Figure 3. A maximum concentration of 6023 ppm is observed at FMQ-2. The Cr concentration in the glass drops to 4095 ppm at FMQ and 3260 ppm at FMQ+2. The value at FMQ-2 is thus similar to that of the MORB melt, but values do not decrease to the same extent with an increase in $fO_2$.

### 3.3. $Cr_2O_3$-MORB melt system

After dissolution of chromium oxide ($Cr_2O_3$) in the MORB melt, a chromite was formed. At low oxygen fugacity (FMQ-2), crystals of $Cr_2O_3$ were completely transformed to chromite. At higher oxygen fugacity (FMQ and FMQ+2), crystals are zoned with residual chromium oxide surrounded by rims of chromite with average thickness ~40 μm. A spongy texture was not observed. At FMQ-2, chromite contains an average of 61.2 wt% of $Cr_2O_3$ in the central zone of the crystals and 58.7 wt% of $Cr_2O_3$ at the rim. At the FMQ buffer the rim chromite has 66.6 wt% of $Cr_2O_3$, while at ΔFMQ+2, rim chromite has an average of 63.2 wt% $Cr_2O_3$.

Magnesium content in the chromite is 13 wt% of MgO at ΔFMQ-2, whereas it is 10.5 wt% at FMQ and 10.9 wt% at FMQ+2 in the chromite rims. Titanium oxide content in the chromite varies from 0.34 to 0.41 wt%, which is close to values in the system of chromite-MORB (0.36-0.41 wt% of $TiO_2$). Total iron oxide concentration in the chromite changes with oxygen fugacity increasing with increasing $f_{O2}$ (13.3 wt% at FMQ-2, 16.5 wt% at FMQ, 18.6 wt% at FMQ+2 of FeO). $Al_2O_3$ concentrations are highly variable, both between and within experiments. For example, at FMQ-2 $Al_2O_3$ concentration is 11.7±0.5 wt%, whereas it is 4.2±3.3 and 5.1±3.1 wt% at FMQ and FMQ+2, respectively.

$Al_2O_3$ varies from 15 to 16 wt% for all oxygen fugacities and it is similar to the initial 15 wt% value. Total iron oxide (FeO) content is a function of oxygen fugacity (6.3 wt% at FMQ-2, 8.5 wt% at FMQ and 9.7 wt% at FMQ+2). Magnesium (MgO) content of 8 wt% is constant and does not change with $f_{O2}$. $TiO_2$ ranges from 1.54 to 1.63 wt%, which is slightly higher than the initial (1.5 wt%) content. Chromium oxide content, as illustrated in Figure 3, reaches a maximum of 6415 ppm at the most reduced conditions (FMQ-2). The Cr content drops to 3058 ppm at FMQ and 1948 ppm at FMQ+2 with increasing oxygen fugacity.

## 3.4. Cr$_2$O$_3$ - haplobasalt system

Chromium oxide dissolved in haplobasaltic melt (Di$_{63}$An$_{37}$) was partly replaced by magnesiochromite. When the oxygen fugacity is low (FMQ-2), crystals of Cr$_2$O$_3$ were completely transformed to magnesiochromite with little zonation (71.78 of Cr$_2$O$_3$ in the center and 68.16 at the rim). At more oxidized conditions (FMQ and FMQ+2), residual crystals have a central zone of almost pure Cr$_2$O$_3$ (98 wt%), with rims of magnesiochromite (73.32 wt% of Cr$_2$O$_3$ at FMQ and 73.35 wt% of Cr$_2$O$_3$ at FMQ+2) with a thickness of ~40 μm (**Figure 5**). The rim consists of euhedral crystals, indicating that the magnesiochromite is in equilibrium with the surrounding glass. Magnesia (MgO) content in magnesiochromites is stable and varies slightly from 19.3 to 20.7 wt%. Titanium oxide is quite low in the magnesiochromite (at the level of the detection limit of 0.15 wt%). Al$_2$O$_3$ in the magnesiochromite varies from 3.1 to 7.7 wt%.

Aluminium oxide content in the glasses is around 8.5 wt% for all three experiments. This value is lower than the original (15 wt% of Al$_2$O$_3$). Magnesia is close to 14.5 wt% for all three glasses that is higher than starting 11.3 wt% content. As shown in **Figure 3**, the highest concentrations of chromium were found in the Cr$_2$O$_3$ – haplobasalt system. The maximum value of 6824 ppm Cr corresponds to the system saturation with magnesiochromite at FMQ-2. The concentration of chromium in the glass decreases when conditions become more oxidized. At FMQ the chromium content is 4498 ppm, furthermore, at FMQ+2, the Cr concentration is 3500 ppm. The initial system is iron free, but iron has been detected in the final products. The total iron oxide content in crystals varies from the detection limit to 2.72 wt%. In the glass, FeO varies from detection limit to 1.59 wt%.

## 4. Model of chromite solubility in mafic and ultramafic melts

Based on our new data on magnesiochromite and chromite solubility and previous data on chromite solubility in the mafic and ultramafic melts, we are able to develop a new predictive model of Cr solubility in mafic and ultramafic melts. The framework for this model is based upon the oxidation of chromium (II) to chromium (III) in the melt described by the reaction:

$$Cr^{2+}_{liq} + \frac{1}{4} O_2 = Cr^{3+}_{liq} + \frac{1}{2}O^{2-}_{liq} \qquad \text{(eqn. 1)}.$$

The equilibrium constant of this reaction is:

$$K = \frac{a_{Cr3+}^{liq}}{a_{Cr2+}^{liq}} * \frac{1}{f_{O_2}^{1/4}}, \quad (eqn.\ 2),$$

where K is the equilibrium constant, $a_{Cr3+}^{liq}$ is activity of the $Cr^{3+}$ in the liquid and $a_{Cr2+}^{liq}$ is activity of $Cr^{2+}$ in the liquid and $a_{O2-}^{liq\ 1/2} = 1$ is activity of $O^{2-}$ in the silicate liquids. It is known that at equilibrium:

$$\Delta G = \Delta G° + R\ T\ \ln K = 0, \quad (eqn.\ 3),$$

where ΔG is the change in Gibbs free energy, ΔG° is the standard Gibbs free energy change, T is temperature in Kelvin, and R is the gas constant (in $J*K^{-1}*mol^{-1}$). Equations 2 and 3 may thus be combined to give:

$$\Delta G° = -R\ T\ \ln\left(\frac{a_{Cr3+}^{liq}}{a_{Cr2+}^{liq}} * \frac{1}{f_{O_2}^{1/4}}\right) \quad (eqn.\ 4),$$

which may be rewritten as:

$$e^{\frac{-\Delta G^0}{RT}} = \frac{a_{Cr3+}^{liq}}{a_{Cr2+}^{liq}} * \frac{1}{f_{O_2}^{1/4}} \quad (eqn.\ 5).$$

At equilibrium, the activity of $Cr^{3+}$ in the chromite (Chr) should be equal to the activity of $Cr^{3+}$ in the liquid such that:

$$a_{Cr3+}^{liq} = a_{Cr3+}^{Chr} \quad (eqn.\ 6),$$

Thus:

$$e^{\frac{-\Delta G^0}{RT}} = \frac{a_{Cr3+}^{Chr}}{a_{Cr2+}^{liq}} * \frac{1}{f_{O_2}^{1/4}} \quad (eqn.\ 7).$$

Given that activity is a product of mole fraction and activity coefficient (a = X*ɣ) where X is mole fraction and ɣ is activity coefficient, equation 7 may be expanded to:

$$e^{\frac{-\Delta G^0}{RT}} = \frac{X_{Cr3+}^{Chr} * \unicode{611}_{Cr3+}^{Chr}}{X_{Cr2+}^{liq} * \unicode{611}_{Cr2+}^{liq}} * f_{O_2}^{-\frac{1}{4}} \quad (eqn.\ 8).$$

Since the mole fraction of total chromium in the liquid is the sum of the mole fractions of $Cr^{2+}$ and $Cr^{3+}$ in the liquid:

$$X_{Cr^{tot}}^{liq} = X_{Cr2+}^{liq} + X_{Cr3+}^{liq}, \quad (eqn.\ 9),$$

where $X_{Cr2+}^{liq}$ and $X_{Cr3+}^{liq}$ are molar fractions of $Cr^{2+}$ and $Cr^{3+}$ in the liquid.

At the same time, given that the activity of $Cr^{3+}$ is the same in the liquid and chromite:

$$X_{Cr^{3+}}^{liq} * Y_{Cr^{3+}}^{liq} = X_{Cr^{3+}}^{Chr} * Y_{Cr^{3+}}^{Chr} \quad \text{(eqn. 10)},$$

Combining rearranged forms of (eqn. 9) and (eqn. 10), it may be shown that:

$$X_{Cr^{2+}}^{liq} = X_{Cr_{tot}}^{liq} - \frac{X_{Cr^{3+}}^{Chr} * Y_{Cr^{3+}}^{Chr}}{Y_{Cr^{3+}}^{liq}} \quad \text{(eqn. 11a)},$$

$$X_{Cr^{2+}}^{liq} = \frac{X_{Cr_{tot}}^{liq} * Y_{Cr^{3+}}^{liq} - X_{Cr^{3+}}^{Chr} * Y_{Cr^{3+}}^{Chr}}{Y_{Cr^{3+}}^{liq}} \quad \text{(eqn. 11b)}.$$

From (eqn. 8) and (eqn. 11b) the following equation may be obtained:

$$e^{\frac{-\Delta G^0}{RT}} = \frac{X_{Cr^{3+}}^{Chr} * Y_{Cr^{3+}}^{Chr} * f_{O_2}^{-\frac{1}{4}} * Y_{Cr^{3+}}^{liq}}{(X_{Cr_{tot}}^{liq} * Y_{Cr^{3+}}^{liq} - X_{Cr^{3+}}^{Chr} * Y_{Cr^{3+}}^{Chr}) * Y_{Cr^{2+}}^{liq}} \quad \text{(eqn. 12)},$$

Rearranging equation 12 it may be shown that:

$$e^{\frac{-\Delta G^0}{RT}} * Y_{Cr^{2+}}^{liq} * \frac{X_{Cr_{tot}}^{liq}}{X_{Cr^{3+}}^{Chr}} * Y_{Cr^{3+}}^{liq} = e^{\frac{-\Delta G^0}{RT}} * Y_{Cr^{2+}}^{liq} * Y_{Cr^{3+}}^{Chr} + Y_{Cr^{3+}}^{Chr} * f_{O_2}^{-\frac{1}{4}} * Y_{Cr^{3+}}^{liq} \quad \text{(eqn. 13)},$$

Given that $X^{liq}_{Cr\ tot}$ and $X^{Chr}_{Cr3+}$ are both measurable quantities, equation 13 is rearranged such that the ratio of these terms is alone on the left hand side the equation.

$$\frac{X_{Cr_{tot}}^{liq}}{X_{Cr^{3+}}^{Chr}} = \frac{Y_{Cr^{3+}}^{Chr}}{Y_{Cr^{3+}}^{liq}} + \frac{Y_{Cr^{3+}}^{Chr} * f_{O_2}^{-\frac{1}{4}} * Y_{Cr^{3+}}^{liq}}{e^{\frac{-\Delta G^0}{RT}} * Y_{Cr^{2+}}^{liq} * Y_{Cr^{3+}}^{liq}} \quad \text{(eqn. 14a)},$$

$$\frac{X_{Cr_{tot}}^{liq}}{X_{Cr^{3+}}^{Chr}} = \frac{Y_{Cr^{3+}}^{Chr}}{Y_{Cr^{3+}}^{liq}} (1 + f_{O_2}^{-\frac{1}{4}} * e^{\frac{\Delta G^0}{RT}} * \frac{Y_{Cr^{3+}}^{liq}}{Y_{Cr^{2+}}^{liq}}) \quad \text{(eqn. 14b)},$$

For simplicity, in the following we will assume ideality of the chromite solid solution (i.e. $Y_{Cr^{3+}}^{Chr}$ has been set to 1). This is not strictly the case (e.g. Sack and Ghiorso, 1991), but as long as $Y_{Cr^{3+}}^{Chr}$ is constant, our approach remains valid. In this case, equation 14b becomes:

$$\frac{X_{Cr_{tot}}^{liq}}{X_{Cr^{3+}}^{Chr}} = \frac{1}{Y_{Cr^{3+}}^{liq}} (1 + f_{O_2}^{-\frac{1}{4}} * e^{\frac{\Delta G^0}{RT}} * \frac{Y_{Cr^{3+}}^{liq}}{Y_{Cr^{2+}}^{liq}}) \quad \text{(eqn. 15)}.$$

This thermodynamically derived equation thus predicts how the partition coefficient of Cr between liquid and chromite (left hand side of equation 15) should vary as a function of T, $f_{O_2}$ and composition. However, the activity coefficient terms of this equation may themselves be a function of temperature and will almost certainly be a function of melt composition. We thus need to consider in a little more detail how each term of equation 15 will vary in order to derive an equation that can be fitted to experimental data with a minimum number of variable parameters.

For example, $\frac{1}{\gamma_{Cr^{3+}}^{liq}}$ is the limit of $\frac{X_{Cr tot}^{liq}}{X_{Cr^{3+}}^{Chr}}$ at elevated $fO_2$, where stability of $Cr^{3+}$ is assured. The experimental data suggest that this term is a sensitive function of composition and temperature (T). We postulate that $\ln \frac{1}{\gamma_{Cr^{3+}}^{liq}}$ can be expressed as a linear function of 1/T and that the parameter "optical basicity" ($\lambda$, Duffy, 1993) can be used as a proxy for melt composition. This choice is made based upon the observations of Berry et al. (2006) described below:

$$\ln \frac{1}{\gamma_{Cr^{3+}}^{liq}} = a + b * \lambda + c/T \qquad \text{(eqn. 16)}.$$

$e^{\frac{\Delta G^0}{RT}}$ is a term which is a function of temperature (of form $\frac{\Delta G^0}{RT} = a_2 + [k/T]$), but is independent of melt composition. The term $\frac{\gamma_{Cr^{3+}}^{liq}}{\gamma_{Cr^{2+}}^{liq}}$ is assumed to be a function of melt composition related to optical basicity, as demonstrated by Berry et al. (2006). Therefore, $\ln \frac{\gamma_{Cr^{3+}}^{liq}}{\gamma_{Cr^{2+}}^{liq}}$ can be expressed in a linear form:

$$\ln\left(\frac{\gamma_{Cr^{3+}}^{liq}}{\gamma_{Cr^{2+}}^{liq}}\right) = a_3 + g * \lambda, \qquad \text{(eqn. 17)},$$

where $a_3$ and g are constants. Combining these expressions into equation (15) results in:

$$\frac{X_{Cr tot}^{liq}}{X_{Cr^{3+}}^{Chr}} = \frac{X_{Cr tot}^{liq}}{X_{Cr_2O_3}^{Chr}} = exp(a + b * \lambda + c/T) * (1 + f_{o2}^{-\frac{1}{4}} * exp\left(a_2 + [\frac{k}{T}]\right) * exp(a_3 + g * \lambda) \quad \text{(eqn. 18)}.$$

Parameters $a_2$ and $a_3$ cannot be derived independently and have been combined to a single constant, d (=$a_2+a_3$), giving:

$$\frac{X_{Cr tot}^{liq}}{X_{Cr_2O_3}^{Chr}} = exp(a + b * \lambda + c/T) * (1 + f_{o2}^{-\frac{1}{4}} * exp(d) * exp\left(\left[\frac{k}{T}\right]\right) * exp(g * \lambda)) \qquad \text{(eqn. 19a)}.$$

or:

$$\frac{X_{Cr.tot}^{liq}}{X_{Cr_2O_3}^{Chr}} = exp(a + b*\lambda + c/T) * (1 + f_{o2}^{-\frac{1}{4}} * exp(d + k/T + g*\lambda)) \quad \text{(eqn. 19b)},$$

where optical basicity ($\lambda$) was calculated based on the work of Duffy (1989; 1993):

$$\lambda = \sum \frac{\lambda ox * Cox * No *}{Mox * \sum \frac{Cox * No}{Mox}}, \quad \text{(eqn. 20)},$$

where $\lambda_{ox}$ is the optical basicity of each oxide (taken from Table 2 in Duffy (1993)), $C_{ox}$ is concentration of the oxide in the glass in wt%. *No* is number of oxygens in the formula of oxide. $M_{ox}$ is molar weight of the oxide.

The model was fitted using the experimental data of Murck & Campbell (1986), Barnes (1986), Roeder & Reynolds (1991), Forsythe & Fisk (1994), Poustovetov & Roeder (2000), Voigt et al. (2016) and those of this study. The data were chosen for the pairs of good glass and spinel analysis. For this fitting, we have used experimental data for basaltic melts obtained at 1 bar and the following parameters: temperatures ranging from 1153 °C to 1450 °C; logf$_{O2}$ ranging from -12.8 to -0.7; and λ from 0.56 to 0.61 (Electronic Appendix). Coefficients for equation 19b were found using the "curve_fit" function of Python programming language (see Appendix 1). Python libraries such as NumPy, Pandas, SymPy, SciPy, and Matplotlib were employed. The final coefficients for equation 19b are: a (-7.01 ± 2.10), b (13.72 ± 2.73), c (-12405 ± 1253), d (24.46 ± 3.13), k (-24395 ± 2037), g (-23.59 ± 4.20). **Figure 6** illustrates $\frac{X_{Cr.tot}^{liq}}{X_{Cr_2O_3}^{Chr}}$ calculated from the model (eqn. 19b) as a function of $\frac{X_{Cr.tot}^{liq}}{X_{Cr_2O_3}^{Chr}}$ measured from our own experiments or from the literature sources. Comparison of measured and predicted values has a slope of 0.9425 (rather than 1 in the ideal case) and an intercept of 0.0001 (rather than 0 in the ideal case). The correlation coefficient $R^2$ is 0.94. Overall, we thus conclude that the model has a level of agreement that is excellent.

## 5. Discussion

**Figure 7** depicts chromium content in basaltic glasses in equilibrium with chromites as a function of oxygen fugacity and temperature (this study and literature data). This figure clearly illustrates that chromium content in the chromite-saturated melts increases when oxygen fugacity decreases. In line with previous interpretations, we assign this behavior to a combination of a higher $Cr^{2+}$ fraction in the liquid at reducing conditions (Berry et al., 2004, 2006) and a higher solubility of $Cr^{2+}$ compared to $Cr^{3+}$. Higher solubility of $Cr^{2+}$ can be explained by its larger ionic radius and lower charge.

It is clear from **Figure 7** that increasing temperature also favors high concentrations of Cr at chromite/magnesiochromite saturation. Our new data are thus in perfect accordance with the previous literature data on chromite (Fe-bearing) solubility. Our point for chromite-MORB system at 1440 °C and FMQ buffer is very close to the point of chromite dissolution in basalt 401 at FMQ and 1450 °C of Murck & Campbell (1986).

Moreover, at FMQ+2, the Cr concentrations are higher for Fe-free system with higher optical basicity (0.62) compared to the Fe-bearing system with lower optical basicity (0.58). This is in contrast to FMQ-2, where the solubility of all four systems is quite close to each other. Thus, at high oxygen fugacity the influence of the optical basicity is very important, but at lower oxygen fugacity this effect becomes negligible. At high $f_{O2}$ the values of $\frac{X^{liq}_{Cr_{tot}}}{X^{Chr}_{Cr_2O_3}}$ converge to the $\frac{1}{\gamma^{liq}_{Cr^{3+}}}$ term such that the observed compositional dependence must be related to that term. At lower $f_{O2}$, $\frac{X^{liq}_{Cr_{tot}}}{X^{Chr}_{Cr_2O_3}}$ is affected by two terms of equation 19b. Given that the constants *b* and *g* are opposite in sign this explains why a reduction of $f_{O2}$ leads to attenuation of the compositional effect.

In addition, Roeder & Reynolds (1991) noted that higher iron content in the liquid corresponds to lower concentration of chromium in the liquid. They explained this assuming that $Cr^{2+}$ and $Cr^{3+}$ hold the same structural positions in the melt as $Fe^{2+}$ and $Fe^{3+}$, respectively. In this study, magnesiochromite was formed in the iron-free ($Cr_2O_3$-haplobasalt) system according to the reaction $Cr_2O_{3solid} + MgO_{liq} = MgCr_2O_{4solid}$. The equilibrium melt formed in this system has the highest chromium concentration of the mafic melts studied. This result may have important implications for natural systems as discussed below.

To estimate silicate liquid composition effect on solubility of chromite, optical basicity was calculated for range of compositions from picritic to rhyolitic from GeoRoc (http://georoc.mpch-mainz.gwdg.de/georoc/) and PetDB (http://www.petdb.org) databases. The distribution coefficients $\frac{X^{liq}_{Cr_{tot}}}{X^{Chr}_{Cr_2O_3}}$ were calculated for this range of optical basicity at fixed temperature of 1200 °C and an oxygen fugacity corresponding to the FMQ buffer according to the eqn. 19b. $\text{Log}\frac{X^{liq}_{Cr_{tot}}}{X^{Chr}_{Cr_2O_3}}$ is plotted as function of optical basicity in **Figure 8**. It can be seen that the function has a minimum, due to the fact that there is competition between the effect of melt composition on the Cr content at high $f_{O_2}$ and the highest $Cr^{2+}/Cr^{3+}$ ratios (equation 19b). However, overall, solubility typically increases from felsic to ultramafic melts.

To study the temperature effect, the distribution coefficients $\dfrac{X^{liq}_{Cr^{tot}}}{X^{Chr}_{Cr_2O_3}}$ were calculated according to eqn. 19b for a range of temperature from 900 to 1500 °C at fixed log $f_{O_2}$ = -8.4 and optical basicity = 0.588 (corresponding to the optical basicity of the initial basalt in our experiments: MORB 3786/3). As shown in **Figure 9**, temperature strongly increases the solubility of chromite in the silicate melt.

Finally, the dependence of Cr distribution coefficient $\dfrac{X^{liq}_{Cr^{tot}}}{X^{Chr}_{Cr_2O_3}}$ on oxygen fugacity is illustrated in **Figure 10**. The coefficients were calculated according to eqn. 19b for range of log $f_{O_2}$ from -13.0 to -0.1 at fixed temperature of 1200°C for basalt with optical basicity = 0.588 (starting MORB 3786/3). It can be seen that solubility decreases asymptotically with an increase of log $f_{O_2}$. Thus, our model reproduces the oxygen fugacity effect found in all experiments (**Fig. 7**).

Comparing the effects of individual parameters, we find that the range of log $\dfrac{X^{liq}_{Cr^{tot}}}{X^{Chr}_{Cr_2O_3}}$ caused by a change of optical basicity corresponding to a variation from picrite to rhyolite (-3.2 to -2.9, **Figure 8**) represent change of temperature of less than 140°C or a change of oxygen fugacity of 6.3 logarithmic units, the latter range being large because of the fact that values of log $\dfrac{X^{liq}_{Cr^{tot}}}{X^{Chr}_{Cr_2O_3}}$ flatten off at high $f_{O_2}$ when there is little or no $Cr^{2+}$. When taken together, **Figures 8 to 10** indicate that temperature has the most significant influence on solubility of chromite, followed by oxygen fugacity at low $f_{O_2}$ then melt composition and oxygen fugacity at high $f_{O_2}$. Considering temperature and $f_{O_2}$ in a little more detail, these two parameters were considered independently in the discussion above. However, geological systems tend to evolve along paths that are parallel to solid buffers such as FMQ. For this reason, we have also predicted the variation of $\dfrac{X^{liq}_{Cr^{tot}}}{X^{Chr}_{Cr_2O_3}}$ along IW (iron-wustite buffer), FMQ-2, FMQ, FMQ+2 mineral buffers in the range of temperatures from 900 to 1500 °C for MORB 3786/3 with optical basicity 0.588. Values of $\dfrac{X^{liq}_{Cr^{tot}}}{X^{Chr}_{Cr_2O_3}}$ calculated from eqn. 19b were then multiplied by the molar fraction of Cr in typical chromite (50 wt% of $Cr_2O_3$: $X_{Cr_2O_3}^{Chr}$ =0.308) to obtain the mole fraction of the chromium in the liquid, that was then recalculated to concentration in ppm. The logarithm of Cr concentrations in the silicate liquid along IW, FMQ-2, FMQ, FMQ+2 mineral buffers *versus* temperature is illustrated in **Figure 11**. The lines are close to each other at low temperatures, which implies a weak influence of oxygen fugacity, but lines diverge at high temperature. For instance, the difference of Cr concentrations in the silicate melt in equilibrium with chromite derived at 900 °C at different $fO_2$ is insignificant. At 1200 °C, near basaltic liquidus, the difference between Cr concentrations at FMQ-2 (940 ppm Cr) and FMQ (610 ppm Cr) is near 300 ppm. However, at high temperature 1500 °C this difference between FMQ-2 (7680 ppm Cr) and FMQ (3750 ppm Cr) buffers is almost 4000 ppm.

The simultaneous influence of oxygen fugacity and melt composition on chromite solubility at a fixed temperature 1200 °C is illustrated in **Figure 12**. These values were calculated in the same way as for **Figure 11** (i.e. assuming $X_{Cr_2O_3}^{Chr} = 0.308$ and a sum of mole fractions corresponding to that of the starting MORB 3786/3 glass). These assumptions both introduce some uncertainty, but combined uncertainty is estimated to be <±20%. At high oxygen fugacity (**Fig. 12**) an increase in optical basicity (i.e. more mafic liquids) increases the concentration of Cr in chromite saturated liquids. For example, at a log $f_{O2}$ = -2, Cr concentration increases from 250 to 750 ppm. Interestingly, at very low oxygen fugacity the effect of optical basicity is opposite. For example, at log $f_{O2}$ of -12.0, increasing optical basicity from 0.495 to 0.636 (from rhyolite to picrite), Cr concentrations drop from 3250 ppm to 1750 ppm. At log $f_{O2}$ = -8.4 (corresponding to FMQ buffer at 1200 °C) a modest increase from 600 to 1000 ppm is observed during increase of optical basicity along whole range of melt compositions from rhyolite to picrite. Thus, picritic melts saturated with chromite are enriched in Cr at FMQ buffer and 1200 °C, whereas at extremely high oxygen fugacity at the same temperature of 1200 °C, rhyolitic melts would have much lower Cr contents. Therefore, mixing or hybridization of typical basaltic melt with $SiO_2$-rich melts/rocks would result in massive precipitation of chromite from the mafic liquid at FMQ and 1200 °C, that is in excellent accordance with classic model of Irvine (1975) in implications mostly for the layered intrusions.

In summary, compared to the existing models, our model provides dependence of the chromite/melt partitioning on the most important parameters controlling the partitioning: temperature, oxygen fugacity and compositional parameter, which is optical basicity. In contrast, Murck and Campbell's eq.17 ignores the temperature parameter. This is a problem of the model, which does not consider the most important temperature parameter controlling chemistry and crystallization field of chrome spinel. Other parameter, which may be quantitatively estimated and compared to the estimations made by our models, is $X_{Cr_2O_3}$ in chrome spinel. Our model compared to the models of Poustovetov & Roeder (2000) and Nikolaev et al. (2018) gives better results when calculating $X_{Cr_2O_3}$ in chromite (**Fig. 13a,b**). It may be seen that the accordance of the calculated $X_{Cr_2O_3}$ to the experimental ones at co-existing equilibrium chromite is much better with our model. Additionally, our model expands the compositional field of the model application to felsic and intermediate melts never investigated experimentally.

## 6. Implications

If a mantle-derived magma is initially reduced, but starts to degas and become more oxidized, the chromite would be crystallized. **Figure 10** demonstrates that stability of chromite may be caused by the mafic magma oxidation in the initially water-rich system. Indeed, the chromitite formation in the oceanic mantle-crust transition zone due to serpentinite dehydration, reaction with basaltic melt and assimilation (Borisova et al., 2012; 2020a,b) corresponds to such natural examples and settings, where initially reduced and hydrous magmas (enriched in $CH_4$ and $H_2$) would degas and progressively oxidize. The massive chromite crystallization is responsible for the chromitite formation at the mantle-crust transition zone. Future model of chromitite formation will be developed in details; it is beyond of our stydy. Additionally, our results show that iron poor/Mg-rich melt systems are favorable for high chromite dissolution, at high temperature, at reducing conditions, and in equilibrium with Cr-rich chromites, all parameters that increase Cr content in silicate liquid. Given that serpentinized mantle is enriched in Mg and that magnesiochromite is the Mg-Cr-rich chromite mineral frequently presents in the serpentinized rocks (e.g. Borisova at al. 2012), assimilation of serpentinized lithosphere by mafic/ultramafic melts would be a viable way to reach high concentrations of Cr in the melt. Fractional crystallization of olivine from these chromium-rich melts would decrease magnesium index (Mg/(Mg+Fe)) of the residual melts, the iron concentration in the melts would be increased accordingly. This effect, combined with the cooling and degassing, may be an efficient mechanism to trigger massive chromite crystallization from hybrid mafic/ultramafic melts. Our model also expands the field of the chromite-saturated melts, mostly of mafic and ultramafic compositions, to intermediate and felsic melts and predicts that the melt crystallization is a viable mechanism to trigger massive chromite crystallization. Indeed, our experiments demonstrate Cr concentration ranges from 225 to 580 ppm at $f_{O_2}$ above

FMQ in experimental felsic to intermediate melts formed upon the hydrated peridotite melting, in particular at the mantle-crust transition zone, and reaction with the basaltic melts (Borisova et al., 2020a,b). These Cr concentrations are in excellent agreement with those predicted by the thermodynamic model constructed in our work (**Fig. 12**). Therefore, the proposed thermodynamic model may be applied to model formation of the oceanic mantle-crust transition zone.

## 6. Conclusions

1. New data on solubility of chromite and magnesiochromite have been obtained for basaltic and haplobasaltic ($\lambda = 0.58$ to $0.6\_$) systems at high temperature (1440 °C) and oxygen fugacity in the range from FMQ-2 to FMQ+2.
2. Magnesiochromite is the equilibrium phase in haplobasatic melt with high Cr contents (3500 – 6800 ppm). This shows high solubility of magnesiochromite in haplobasaltic melt, that has an important implication for serpentinized mantle assimilation by basaltic magmas.
3. A predictive model of chromite and magnesiochromite solubility in silicate melts has been developed. This model is applicable for the explanation of massive chromite crystallization from hybrid mafic/ultramafic melts, happening in the mantle-crust transition zone according to model of disseminated, stratiform and nodular chromitite formation by Borisova et al. (2012), Zagrtdenov et al. (2018) and Rospabé et al. (2019).
4. At moderate to high oxygen fugacities (for example, >FMQ-1 at 1200 °C), chromite solubility increases from felsic to ultramafic melts, but in general terms, temperature is the most important variable. Our model may thus predict physico-chemical conditions of massive crystallization of chromitites. The Cr concentrations in felsic to intermediate melts saturated in chromite and formed upon the hydrated peridotite melting and reaction with basaltic melts (Borisova et al., 2020a,b) are in excellent agreement with those predicted by the thermodynamic model constructed in our work. The formation and involvement of intermediate and felsic melts with high $SiO_2$ contents and low chromite solubility has a direct impact on the formation of the terrestrial chromite-bearing mantle-crust transition zone represented by chromitites and dunites.

# References

Barnes, S.J. (1986). The distribution of chromium among orthopyroxene, spinel and silicate liquid at atmospheric pressure. Geochemica et Cosmochemica Acta 50(9):1,889-1,909. doi:10.1016/0016-7037(86)90246-2.

Berry AJ, O'Neill HSC (2004) A XANES determination of the oxidation state of chromium in silicate glasses. Am Mineral 89:790−798.

Berry AJ, O'Neill HSC, Scott DR, Foran GJ, Shelley J (2006) The effect of composition on $Cr^{2+}/Cr^{3+}$ in silicate melts. Am Mineral 91:1901−1908.

Borisova A.Y., Zagrtdenov N.R., Toplis M.J., Bohrson W.A, Nedelec A., Safonov O.G., Pokrovski G.S., Ceuleneer G., Melnik O.E., Bychkov A.Y., Gurenko A.A., Shcheka S., Terehin A., Polukeev V.M., Varlamov D.A., Chariteiro K.E.A., Gouy S., de Parseval P. (2020a) Planetary felsic crust formation at shallow depth. Frontiers in Earth Science. Special Volume on magma-rock and magma-mush interactions as fundamental processes of magmatic differentiation. (on line).

Borisova A.Y., Zagrtdenov N.R., Toplis M.J., Ceuleneer G., Safonov O., Shcheka S., Bychkov A.Y. (2020b) Serpentinized lithospheric mantle assimilation by oceanic basalts is fast. Frontiers in Earth Science. Special Volume on magma-rock and magma-mush interactions as fundamental processes of magmatic differentiation. (in line).

Borisova A.Y., Ceuleneer G., Arai S., Kamenetsky V., Béjina F., Polvé M., de Parseval Ph., Aigouy T., Pokrovski G.S. (2012a). A new view on the petrogenesis of the Oman ophiolite chromitites from microanalyses of chromite-hosted inclusions Journal of Petrology, 53(12): 2411-2440, doi:10.1093/petrology/egs054.

Borisova A. Y., Toutain J.-P., Stefansson A., Gouy S., and de Parseval Ph. (2012b). Processes controlling the 2010 Eyjafjallajökull explosive eruption. Journal of Geophysical Researc, Vol. 117, B05202, doi:10.1029/2012JB009213.

Borisova A. Y., Zagrtdenov N. R., Toplis M. J., Donovan J. J., Llovet X., Asimow P. D., de Parseval Ph., Gouy S. (2018). Secondary fluorescence effects in microbeam analysis and their impacts on geospeedometry and geothermometry. Chemical Geology 490, 22 – 29.

Duffy J.A. (1989). A common optical basicity scale for oxide and fluoride glasses. Journal of Non-Crystalline Solids 109, 35-39.

Duffy J.A. (1993) A review of optical basicity and its applications to oxidic systems. Geochimica et Cosmochimica acta, Vol. 57, pp. 3961-3970.

Forsythe L.M., Fisk M.R. Comparison of experimentally crystallized and natural spinels from LEG-135. Proceedings of the Ocean Drilling Program. Vol. 135 (1994).

Gale Allison, Dalton Colleen A., Langmuir Charles H., Su Yongjun, Schilling Jean-Guy (2013). The mean composition of ocean ridge basalts. Geochemistry, Geophysics, Geosystems. Volume 14, Issue 3, March 2013. Pages 489–518

Hill R. and Roeder P. (1974). The crystallization of spinel from basaltic liquid as a function of oxygen fugacity. The Journal of Geology 82:709–729.

Irvine T. N., (1975) Crystallization sequences in the Muskox intrusion and other layered intrusions-II. Origin of chromitite layers and similar deposits of other magmatic ores. Geochemica et Cosmochemica Acta, 39, 991-1020.

Jochum K.P., Stoll B., Herwig K., Willbold M., Hofmann A.W., Amini M., Aarburg S., Abouchami W., Hellebrand E., Mocek B., Raczek I., Stracke A., Alard O., Bouman C., Becker S., Dücking M., Brätz H., Klemd R., de Bruin D., Canil D., Cornell D., de Hoog C.-J., Dalpé C., Danyushevsky L., Eisenhauer A., Gao Y., Snow J.E., Groschopf N., Günther D., Latkoczy C., Guillong M., Hauri E.H., Höfer H.E., Lahaye Y., Horz K., Jacob D.E., Kasemann S.A., Kent A.J.R., Ludwig T., Zack T., Mason P.R.D., Meixner A., Rosner M., Misawa K., Nash B.P., Pfänder J., Premo W.R., Sun W.D., Tiepolo M., Vannucci R., Vennemann T., Wayne D. and Woodhead J.D. 2006. MPI-DING reference glasses for in situ microanalysis: New reference values for element concentrations and isotope ratios. Geochemistry, Geophysics, Geosystems, 7 / 2, doi: 10.1029/2005GC001060.

Jochum, K.P. et al. Determination of Reference Values for NIST SRM 610-617 Glasses Following ISO Guidelines. Geostandards and Geoanalytical Research 35, 397 – 429 (2011).

Johan Z., Martin R.F., Ettler V. (2017). Fluids are bound to be involved in the formation of ophiolitic chromite deposits. Eur. J. Mineral. 29, 543 – 555.

John W. Anthony, Richard A. Bideaux, Kenneth W. Bladh, and Monte C. Nichols, Eds., Handbook of Mineralogy, Mineralogical Society of America, Chantilly, VA 20151-1110, USA. http://www.handbookofmineralogy.org/.

Maurel, C. & Maurel, P. (1982a). Etude experimentale de la solubilite du chrome dans les bains silicates basiques et de sa distribution entre liquide et mine raux coexistants Conditions d'existence du spinelle chromifere. Bulletin Mineralogique 105, 640-647.

Maurel, C. & Maurel, P. (1982b). Etude experimentale de la distribution de l'aluminium entre bain silicate basique et spinelle chromifere. Implications petrogenetiques: teneur en chrome des spinelles. Bulletin Mineralogique 105, 197-202.

Murck & Campbell (1986). The effects of temperature, oxygen fugacity and melt composition on the behaviour of chromium in basic and ultrabasic melts. Geochimica et Cosmochimica Acta Vol. 50, pp. 1871-1887.

Nikolaev G. S., Ariskin A. A., Barmina G. S. (2018). SPINMELT-2.0: Simulation of spinel–melt equilibrium in basaltic systems under pressures up to 15 kbar: I. model formulation, calibration, and tests. Geochemistry International, 56, 24–45.

Poustovetov A.A. & Roeder P.L. (2000) The distribution of Cr between basaltic melt and chromian spinel as an oxygen geobarometer. Canadian Mineralogist. Vol. 39, pp. 309-317.

Poustovetov A.A., Roeder P.L. (2001). Contributions to Mineralogy and Petrology 142(1):58-71. Numerical modeling of major element distribution between chromian spinel and basaltic melt, with application to chromian spinel in MORBs. doi: 10.1007/s004100100272.

Roeder, P. L. & Reynolds, I. (1991). Crystallization of chromite and chromium solubility in basaltic melts. Journal of Petrology 32, 909-934.

Rospabé M., Ceuleneer G., Granier N., Arai S., Borisova A.Y. (2019). Multi-scale development of a stratiform chromite ore body at the base of the dunitic mantle-crust transition zone (Maqsad diapir, Oman ophiolite): the role of repeated melt and fluid influxes. Lithos, 350–351, 15, 10523536.

Sack, R. O. & Ghiorso, M. S. (1991). Chromian spinel as petrogenetic indicators: thermodynamics and petrological applications. American Mineralogist 76, 827-847.

Schreiber, H.D. and Haskin, L.A. (1976) Chromium in basalts: Experimental de- termination of redox states and partitioning among synthetic silicate phases. Proceedings of the 7th Lunar Science Conference, p. 1221–1259.

Snow, J. E., Hart, S. R. & Dick, H. J. B. (1994). Nd and Sr isotope evidence linking mid-ocean-ridge basalts and abyssal peridotites. Nature 371, 57-60.

Sushchevskaya, N.M. et al. Magmatism of Mona and Knipovich ridges from spreading zones of polar Atlantic Ocean. Russian Journal of Geosciences. 2, 243 – 267 (2000).

Toplis M.J., Libourel G., Carroll M.R. (1994). The role of phosphorus in crystallisation processes of basalt: An experimental study. Geochimica et Cosmochimica Acta, Volume 58, Issue 2, p. 797-810.

Voigt, Martin, Coogan, Laurence A., von der Handt, Anette, Experimental investigation of the stability of clinopyroxene in mid-ocean ridge basalts: The role of Cr and Ca/Al, LITHOS (2017), doi:10.1016/j.lithos.2017.01.003

Wojtulek, P.M., Puziewicz, J., Ntaflos, T., Bukała, M., 2016. Podiform chromitites from the Variscan ophiolite serpentinites of Lower Silesia (SW Poland) - petrologic and tectonic setting implications. Geological Quarterly 60 (1), 56-66.

Zagrtdenov N., Borisova A. Y., Toplis M., Duployer B.and Tenailleau C. (2015). Preliminary experimental investigation of chromite dissolution in mid-ocean ridge basalt melt. // Goldschmidt Conference Abstracts, 16-21 August 2015, Prague, Cz, Europe.

Zagrtdenov N.R., Ceuleneer G., Rospabé M., Borisova A., Toplis M.J., Benoit M. and Abily B. (2018). Anatomy of a chromitite dyke in the mantle/crust transition zone of the Oman ophiolite. Lithos, 312-313, 343-357.

GeoRoc http://georoc.mpch-mainz.gwdg.de/georoc/

PetDB http://www.petdb.org

# Figures captions

**Figure 1.** Chromium concentrations in natural basaltic and boninitic melts according to the PETDB database (http://www.petdb.org).

**Figure 2.** Residual chromite (Chr) in the basaltic (MORB) glass in the experiment E19. Spongy texture in chromite may be clearly recognized. The image was obtained using an optical microscope in reflected light.

**Figure 3.** Chromium content in the quenched basaltic and haplobasaltic melts in equilibrium with chromite (chromite or magnesiochromite) as a function of oxygen fugacity at the run 1440 °C temperature. Error bars (one standard deviation, characterizing homogeneity of the sample) are smaller than a symbol. Note that the chromite solubility is not strictly linear function of Cr contents on this range of oxygen fugacity.

**Figure 4.** Dependence of distribution coefficients (ratio of $Cr_2O_3$ content in chromite crystals to $Cr_2O_3$ content in the coexisting liquid) on oxygen fugacity at the run 1440 °C temperature.

**Figure 5.** Back-scattered election image of basaltic glass-hosted grain of oxide phases after starting $Cr_2O_3$ dissolution in the haplobasaltic glass ('Di63An37 initially' experiment E33). Residual '$Cr_2O_3$' is observed in the center. Newly formed magnesiochromite ('MgChr') composes euhedral-crystallized rims. It forms euhedral crystals meaning that the magnesiochromite is in equilibrium with the surrounding glass.

**Figure 6.** Fitting of $\frac{X^{liq}_{Cr_{tot}}}{X^{Chr}_{Cr_2O_3}}$ calculated from the model (eqn. 19b) (ordinate) vs. the values calculated based on experiments (abscissa, correlation coefficient is $R^2 = 0.94$).

**Figure 7.** Chromium contents in the mafic and ultramafic quenched melts in equilibrium with chromite and magnesiochromite as function of oxygen fugacity at different temperatures based on this study and literature data. The color dot lines are interpolation of the experimental data obtained at a given temperature at variable oxygen fugacities.

**Figure 8.** Logarithm of Cr distribution coefficient between silicate liquid and chromite in equilibrium with the liquid (expressed as ratio of chromium molar fraction in liquid to $Cr_2O_3$ molar fraction in the chromite) *versus* optical basicity of the liquid at oxygen fugacity corresponding to FMQ mineral buffer and temperature 1200 °C. Range of different effusive rocks from databases GeoRoc and PetDB is taken as examples of silicate melts.

**Figure 9.** Influence of temperature to the chromium distribution coefficient according to eqn. 19b. Section is presented for log $f_{O2}$ = -8.4 and optical basicity (O.B.) = 0.588, corresponding to composition of initial basalt of our experiments (MORB 3786/3).

**Figure 10.** Dependence of Cr distribution coefficient on oxygen fugacity according to eqn. 19b. Section at 1200 °C for basalt with optical basicity = 0.588 (starting MORB 3786/3). Dashed line depicts FMQ buffer at 1200 °C.

**Figure 11.** Logarithm of Cr concentration in the silicate liquid calculated from the eqn. 19b along IW, FMQ-2, FMQ, FMQ+2 mineral buffers vs. temperature for melt with optical basicity = 0.588 (MORB 3786/3).

**Figure 12.** Contour plot of the estimated chromium concentrations in dry silicate melts (in ppm) at chromite saturation as function of optical basicity and logarithm of oxygen fugacity at fixed temperature 1200 °C according to the model (eqn. 19b). Typical ranges of optical basicities for the main silicate melts (rhyolite, andesite, dacite, boninite, basalt and picrite) are marked in brackets at the right side of the plot.

**Figure 13 (a)**. Comparison of our model with the model of Poustovetov & Roeder (2000), which allows calculating $f_{O_2}$ values for given melt saturated with chromite. We demonstrate that new model predicts experimental $X_{Cr_2O_3}$ better than model of Poustovetov & Roeder (2000).

**Figure 13(b).** Comparison of our model with the model of Nikolaev et al. (2018). We demonstrate that new model predicts experimental $X_{Cr_2O_3}$ better than model of Nikolaev et al. (2018).

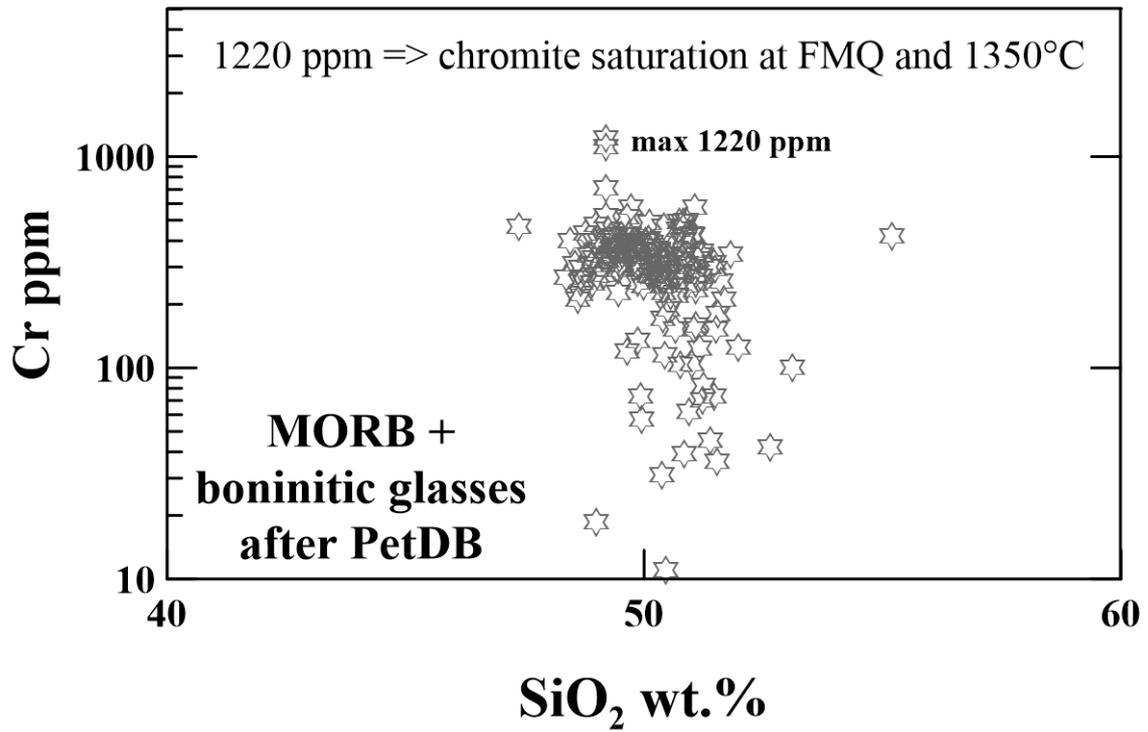

**Figure 1.** Chromium concentrations in natural basaltic and boninitic melts according to the PetDB database (http://www.petdb.org).

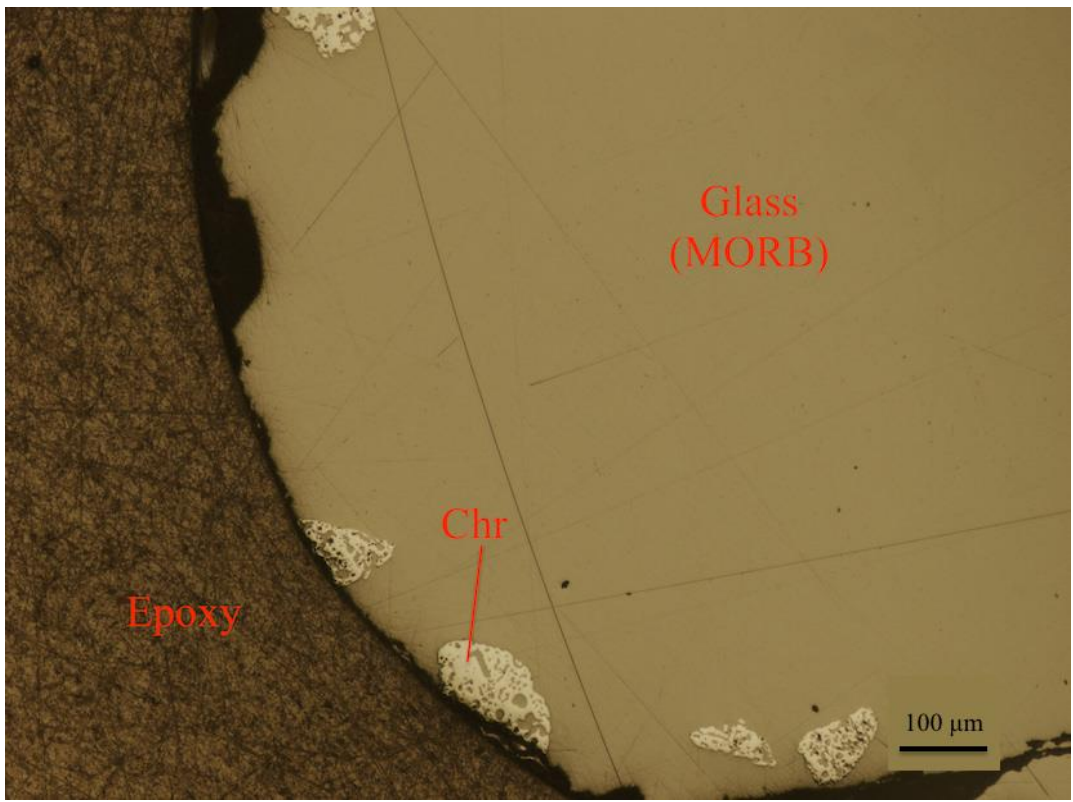

**Figure 2.** Residual chromite (Chr) in the basaltic (MORB) glass in the experiment E19. Spongy texture in chromite may be clearly recognized. The image was obtained using an optical microscope in reflected light.

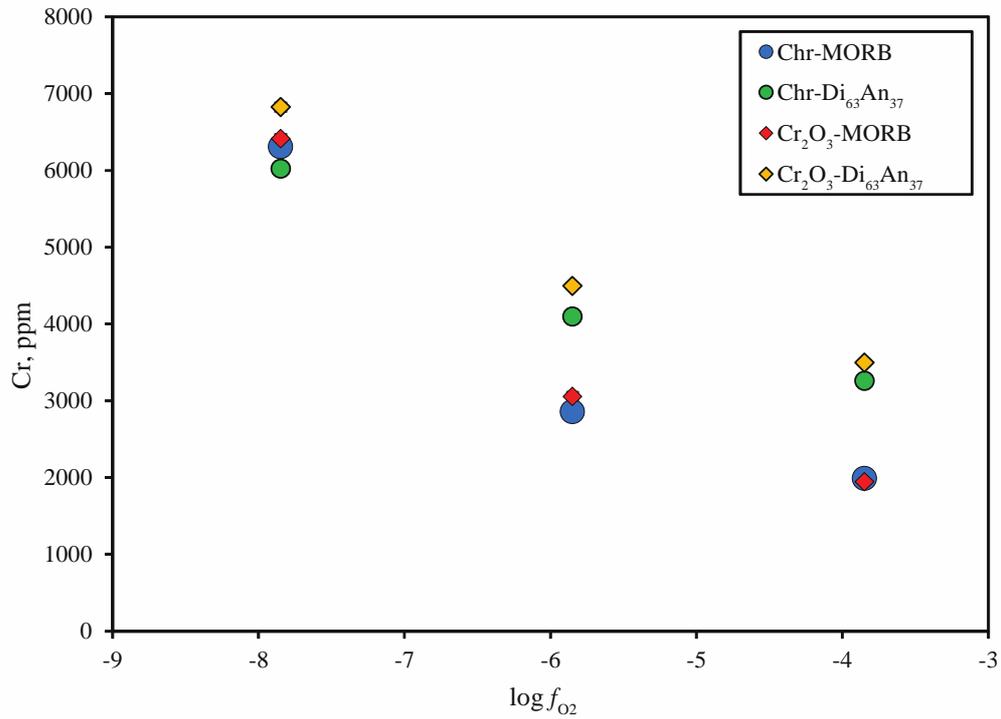

**Figure 3.** Chromium content in the quenched basaltic and haplobasaltic melts in equilibrium with chromite (chromite or magnesiochromite) as a function of oxygen fugacity at the run 1440 °C temperature. Error bars (one standard deviation, characterizing homogeneity of the sample) are smaller than a symbol. Note that the chromite solubility is not strictly linear function of Cr contents on this range of oxygen fugacity.

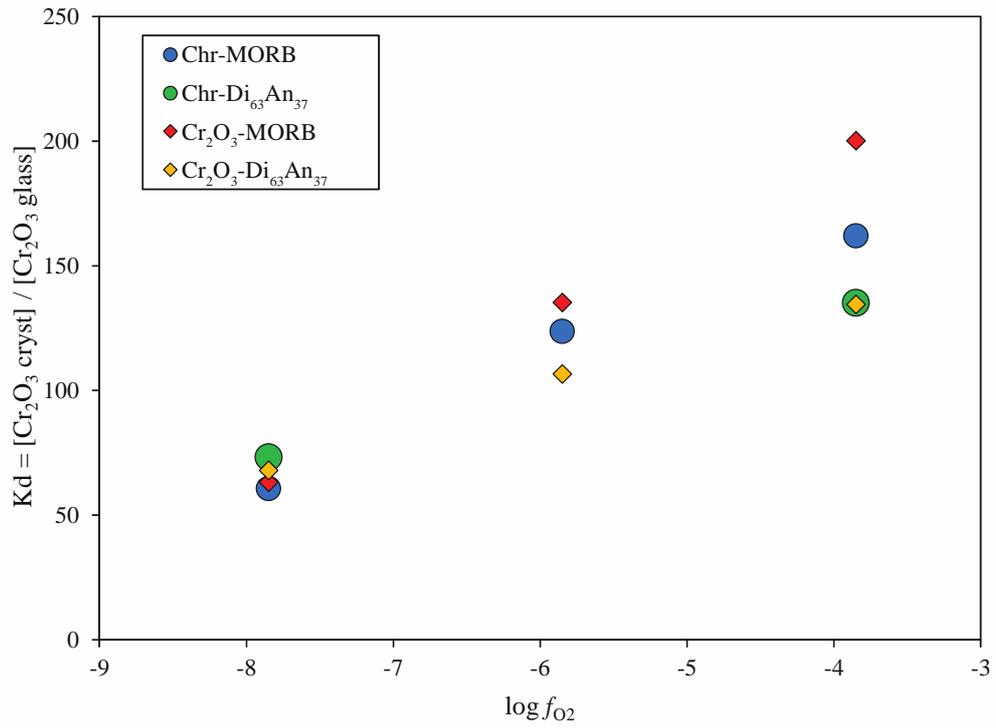

**Figure 4.** Dependence of distribution coefficients (ratio of $Cr_2O_3$ content in chromite crystals to $Cr_2O_3$ content in the coexisting liquid) on oxygen fugacity at the run 1440 °C temperature.

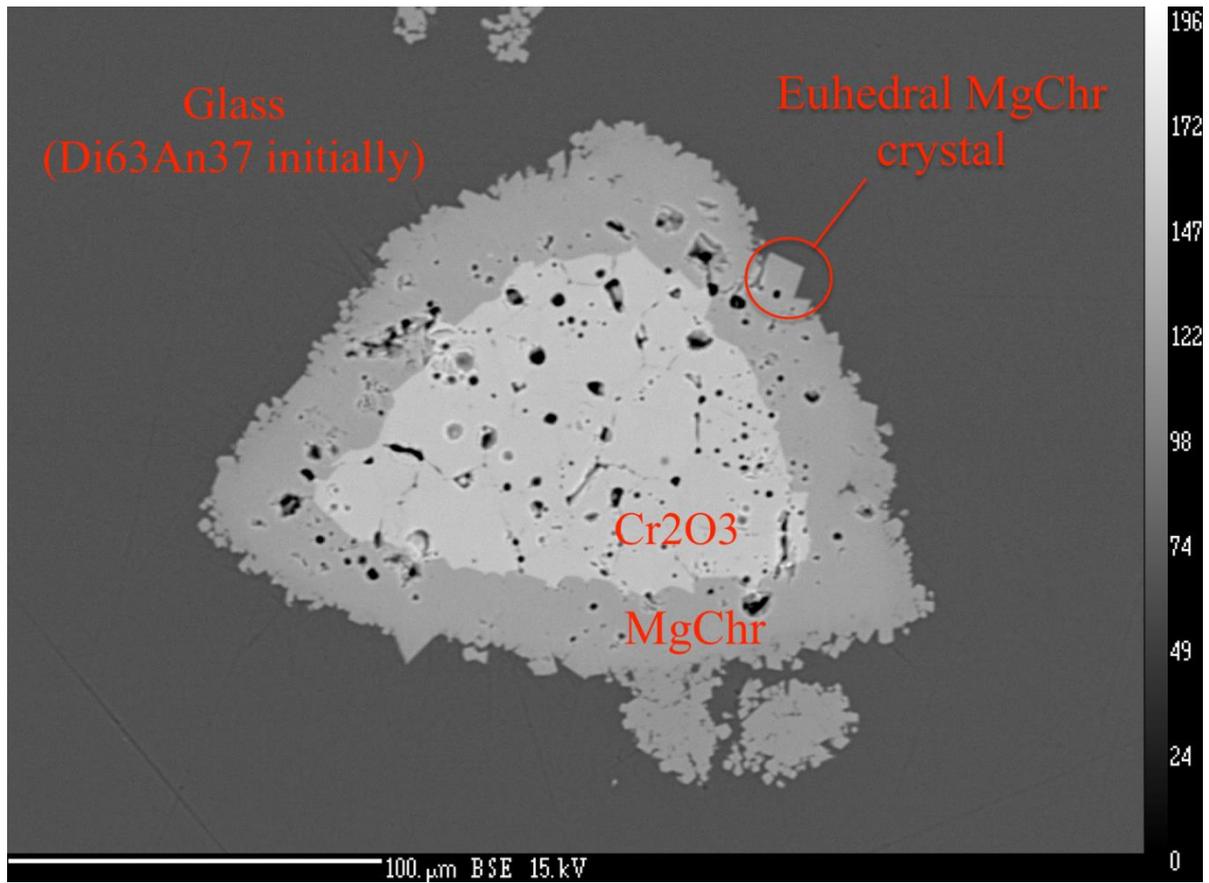

**Figure 5.** Back-scattered election image of basaltic glass-hosted grain of oxide phases after starting $Cr_2O_3$ dissolution in the haplobasaltic glass ('Di63An37 initially' experiment E33). Residual '$Cr_2O_3$' is observed in the center. Newly formed magnesiochromite ('MgChr') composes euhedral-crystallized rims. It forms euhedral crystals meaning that the magnesiochromite is in equilibrium with the surrounding glass.

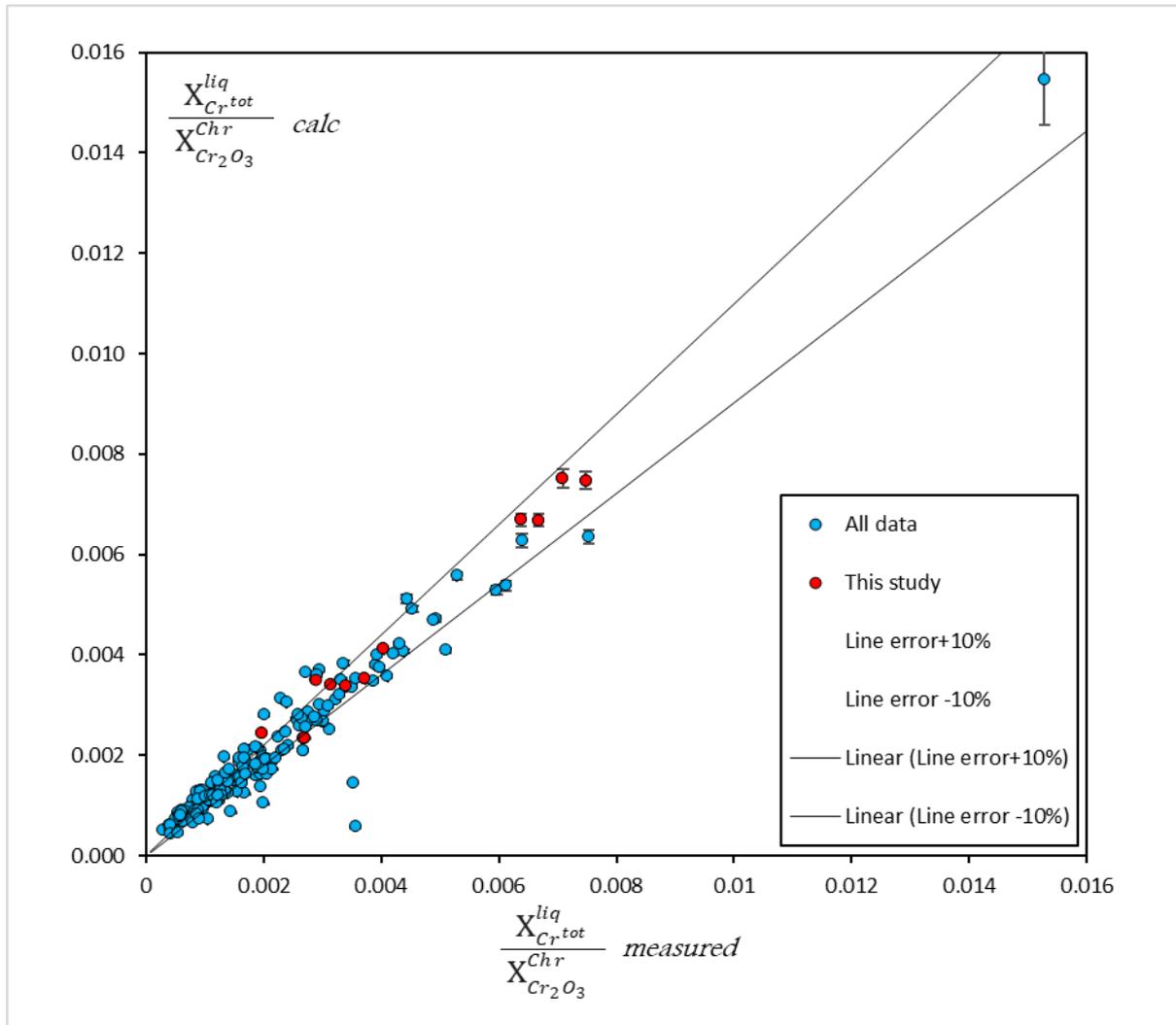

**Figure 6.** Fitting of $\frac{X^{liq}_{Cr^{tot}}}{X^{Chr}_{Cr_2O_3}}$ calculated from the model (eqn. 19b) (ordinate) vs. the values measured based on available experiments (abscissa, correlation coefficient is $R^2 = 0.94$).

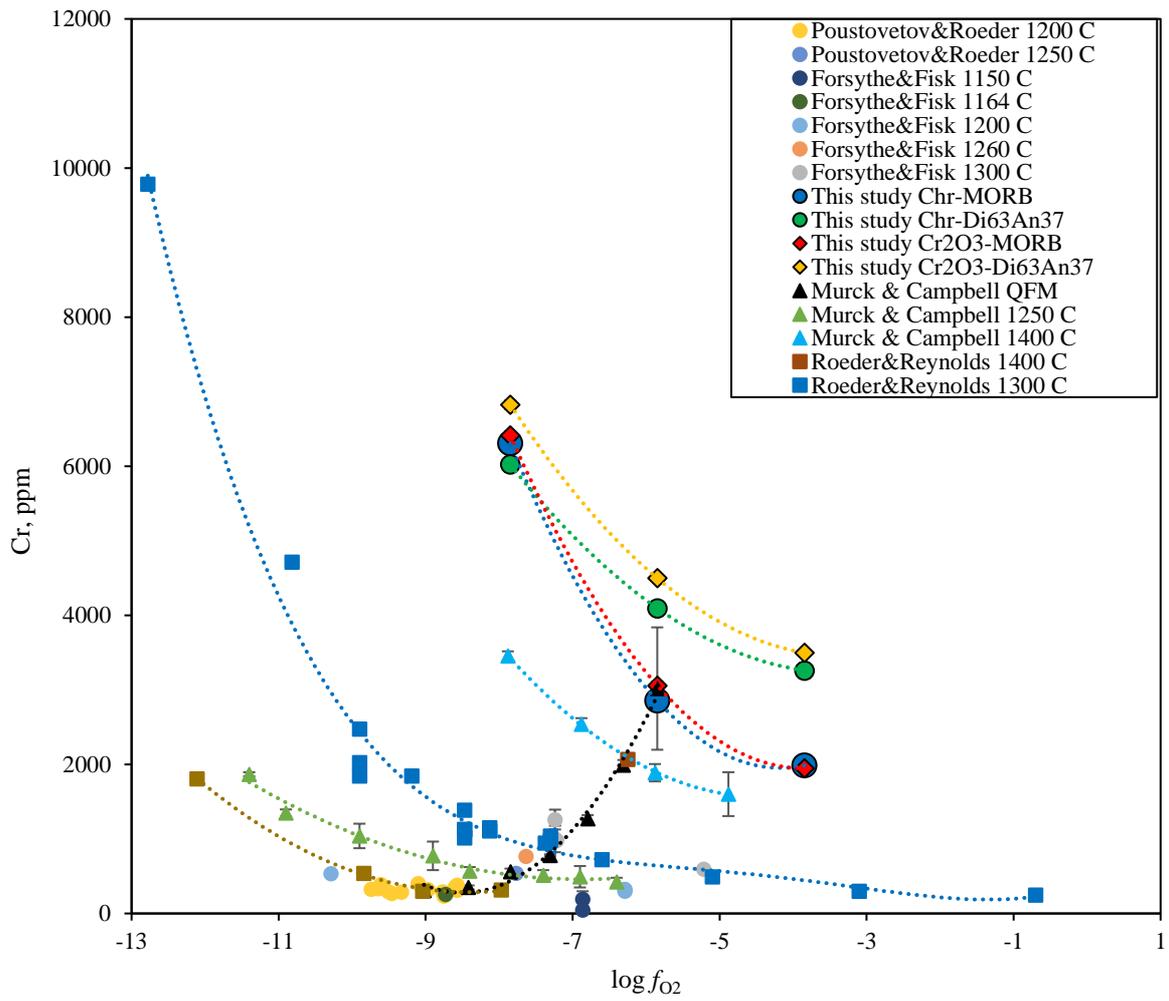

**Figure 7.** Chromium contents in the mafic and ultramafic quenched melts in equilibrium with chromite and magnesiochromite as function of oxygen fugacity at different temperatures based on this study and literature data. The color dot lines are interpolation of the experimental data obtained at a given temperature at variable oxygen fugacities and at a given mineral buffer (FMQ).

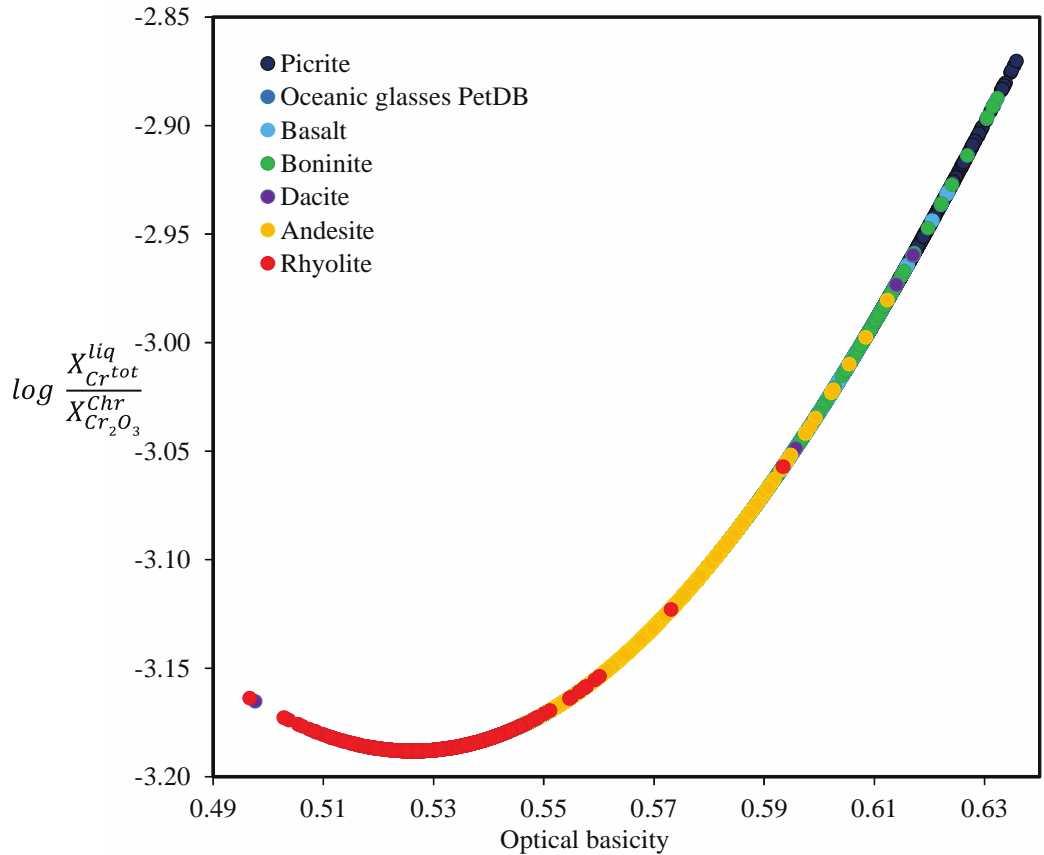

**Figure 8.** Logarithm of Cr distribution coefficient between silicate liquid and chromite in equilibrium with the liquid (expressed as ratio of chromium molar fraction in liquid to $Cr_2O_3$ molar fraction in the chromite) *versus* optical basicity of the liquid at oxygen fugacity corresponding to FMQ mineral buffer and temperature 1200 °C. Range of different effusive rocks from databases GeoRoc (http://georoc.mpch-mainz.gwdg.de/georoc/) and PetDB (http://www.petdb.org) is taken as examples of silicate melts.

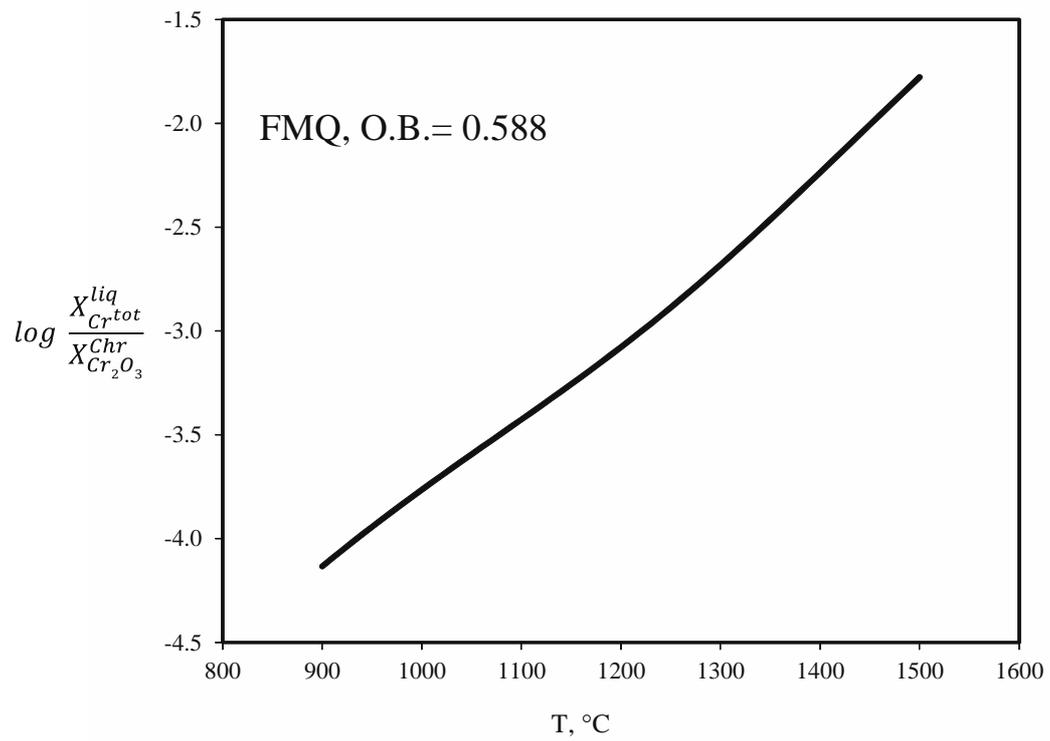

**Figure 9.** Influence of temperature to the chromium distribution coefficient according to eqn. 19b. Section is presented for log $f_{O2}$ = -8.4 and optical basicity (O.B.) = 0.588, corresponding to composition of initial basalt of our experiments (MORB 3786/3).

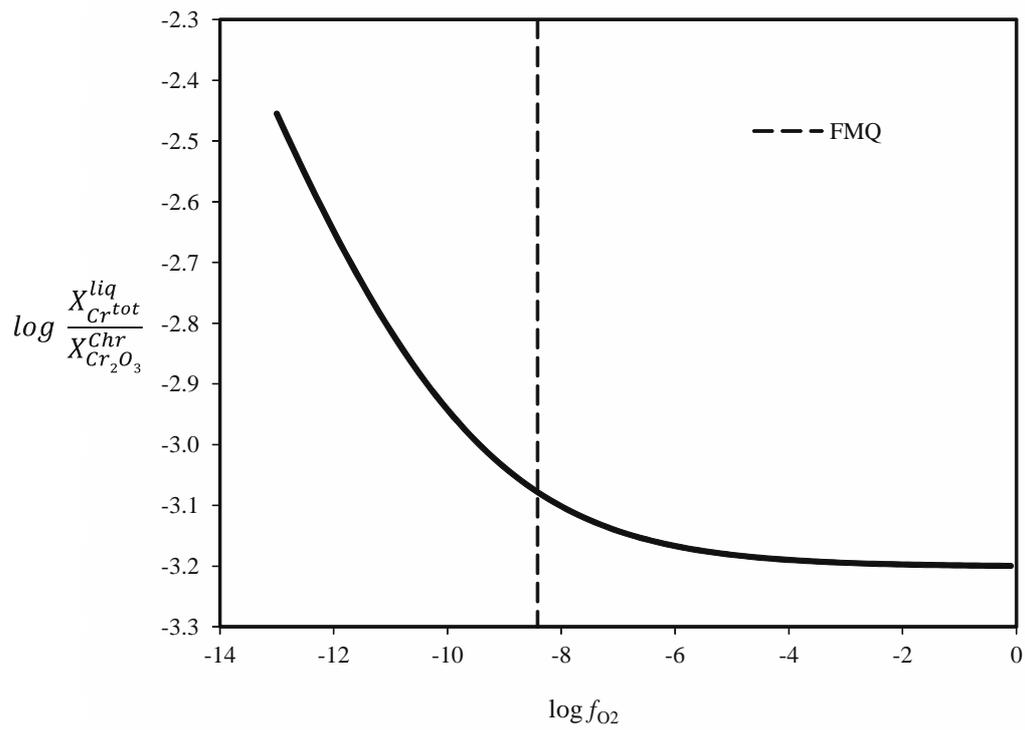

**Figure 10.** Dependence of Cr distribution coefficient on oxygen fugacity according to eqn. 19b. Section at 1200 °C for basalt with optical basicity = 0.588 (starting MORB 3786/3). Dashed line depicts FMQ buffer at 1200 °C.

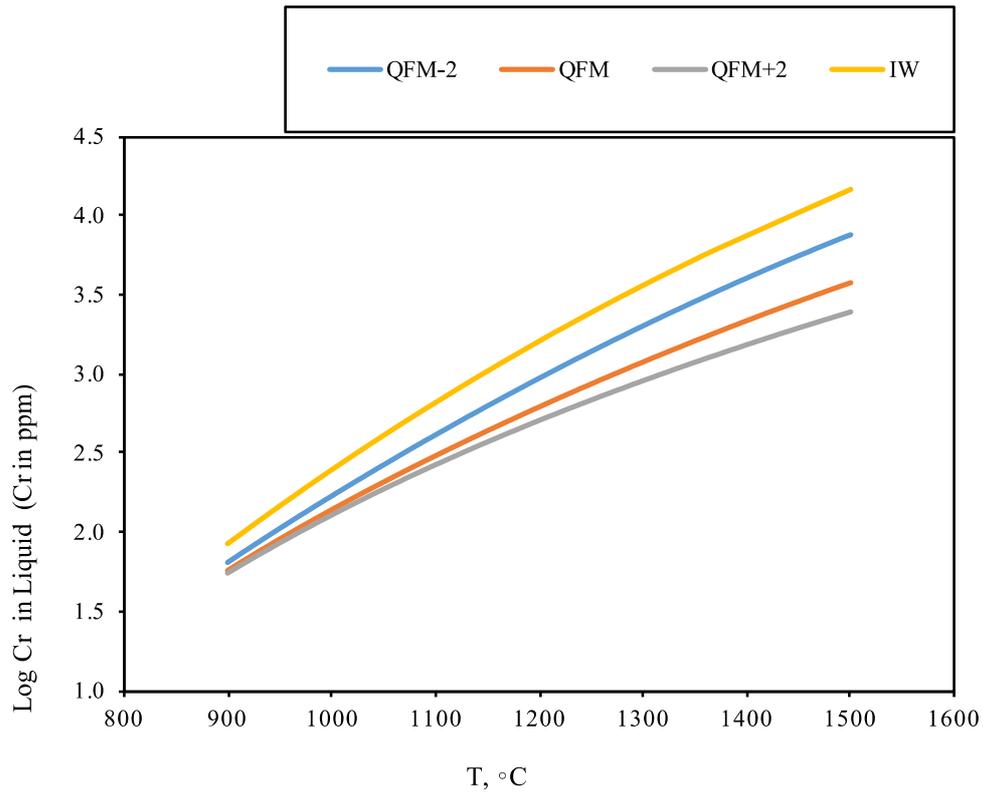

**Figure 11.** Logarithm of Cr concentration in the silicate liquid calculated from the eqn. 19b along IW, FMQ-2, FMQ, FMQ+2 mineral buffers vs. temperature for melt with optical basicity = 0.588 (MORB 3786/3). The combined uncertainty is estimated to be below ±20%.

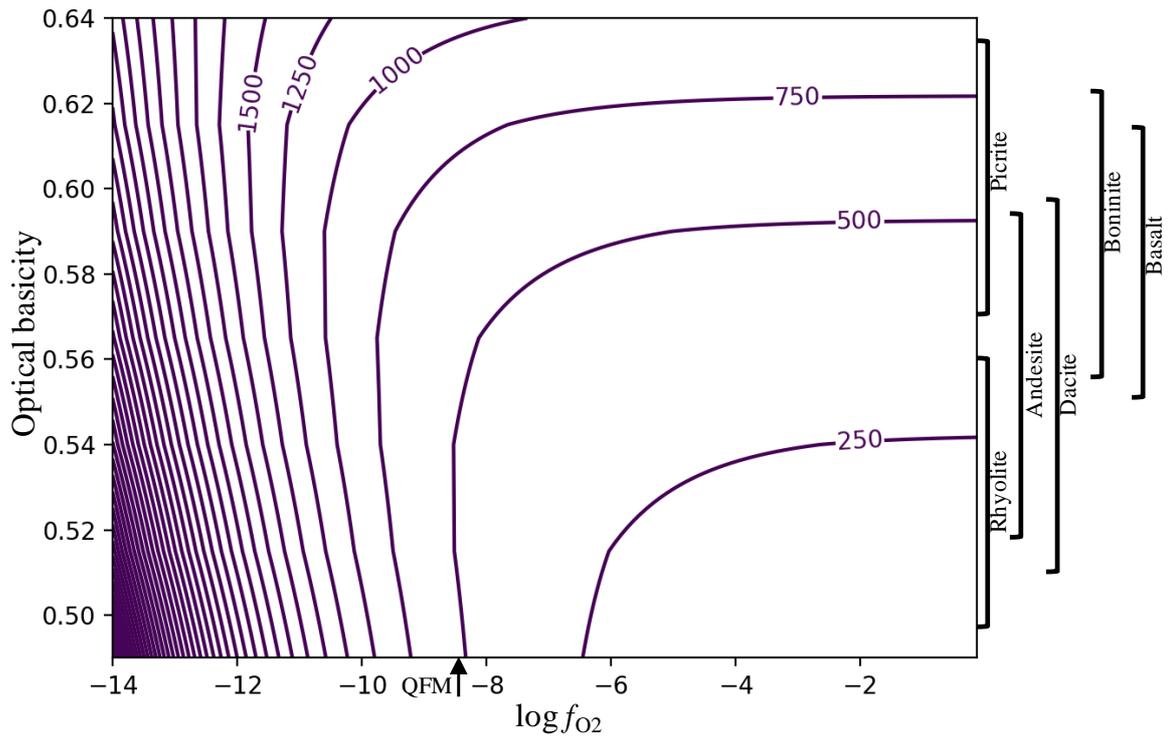

**Figure 12.** Contour plot of the estimated chromium concentrations in dry silicate melts (in ppm) at chromite saturation as function of optical basicity and logarithm of oxygen fugacity at fixed temperature 1200 °C according to the model (eqn. 19b). Typical ranges of optical basicities for the main silicate melts (rhyolite, andesite, dacite, boninite, basalt and picrite) are marked in brackets at the right side of the plot.

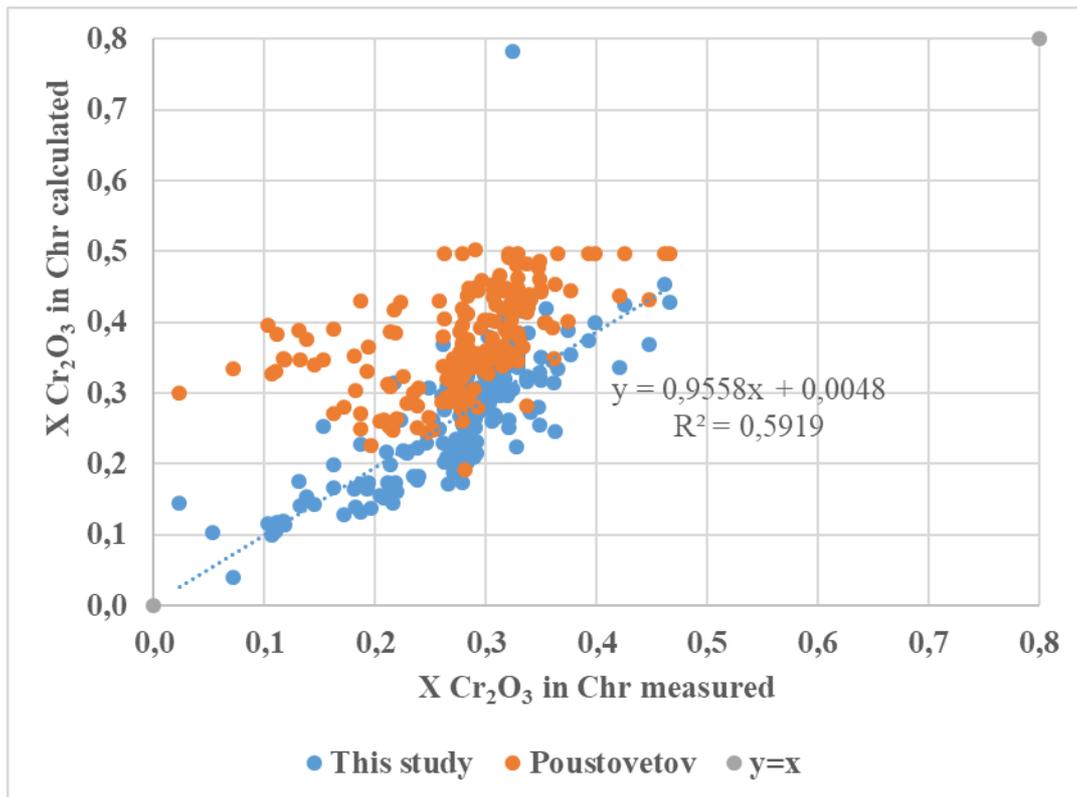

**Figure 13(a).** Comparison of our model with the model of Poustovetov & Roeder (2000), which allows calculating $f_{O2}$ values for given melt saturated with chromite. We demonstrate that new model predicts experimental $XCr_2O_3$ better than model of Poustovetov & Roeder (2000).

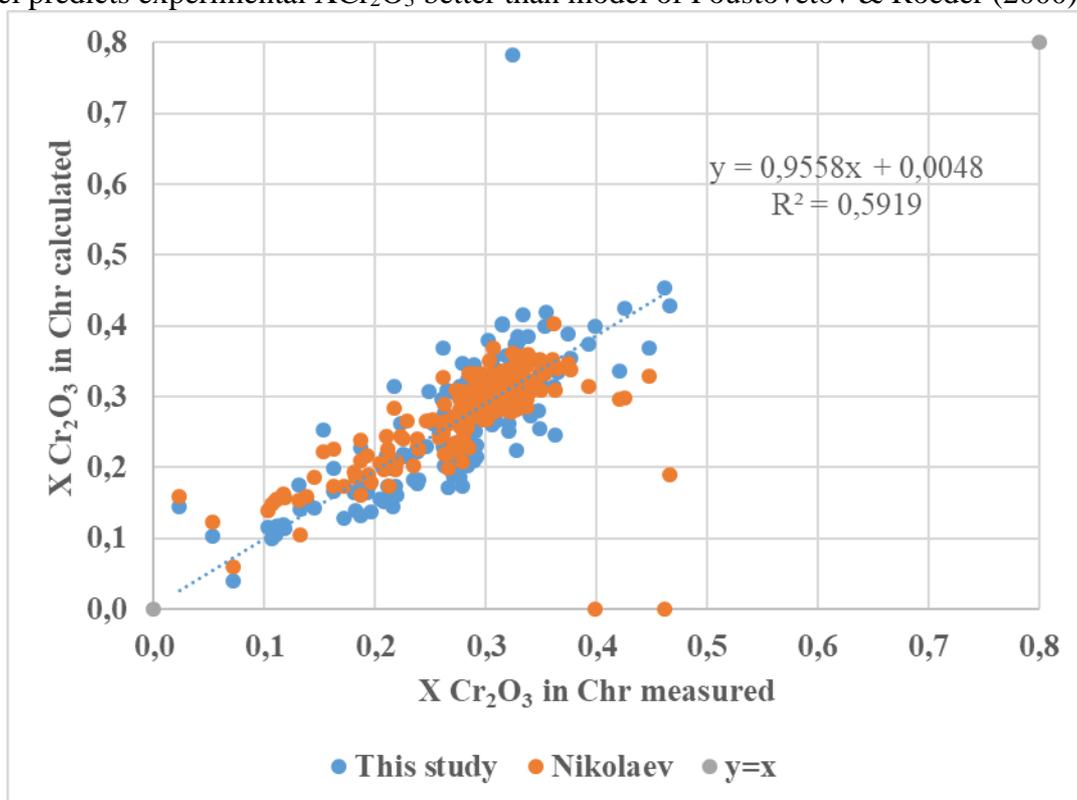

**Figure 13(b).** Comparison of our model with the model of Nikolaev et al. (2018). We demonstrate that new model predicts experimental $XCr_2O_3$ better than model of Nikolaev et al. (2018).

**Table 1. Starting materials compositions**

| | Na$_2$O | SiO$_2$ | Al$_2$O$_3$ | K$_2$O | CaO | MnO | FeO | MgO | TiO$_2$ | P$_2$O$_5$ | Cr$_2$O$_3$ | NiO | Total | Cr(ppm) |
|---|---|---|---|---|---|---|---|---|---|---|---|---|---|---|
| S2 chromite (31)[a] | | | 26.45(42)[b] | | | 0.21(04) | 15.87(62) | 14.81(34) | 0.10(02) | | 41.15(0.27) | 0.16(05) | 98.54 | |
| Cr$_2$O$_3$ (19) | | | 0.14(1) | | | | | 0.28(28) | 1.45(3) | | 97.92(1.19) | | 99.79 | |
| MORB 3786/3[c] | 3.04 | 50.31 | 15.31 | 0.31 | 10.29 | 0.17 | 8.91 | 8.21 | 1.45 | 0.18 | 0.04 | | 98.22 | 275 |
| Di$_{63}$An$_{37}$ (15) | | 50.52(35) | 15.22(25) | | 22.95(31) | | | 11.30(19) | | | | | 99.99 | |

Notes:

[a] Number of analyzes.

[b] Digits in brackets correspond to standard deviation of number of analyzes.

[c] Emission-ICP, relative error <2% for all oxides.

**Table 2. Composition of crystals after experiments and chromium distribution coefficients between crystals and glass (averages from multiple analyses)**

| Run | Rim or Center[a] | System | Buffer | Phase | $Al_2O_3$ | $SiO_2$ | MnO | FeO | MgO | $TiO_2$ | $Cr_2O_3$ | $V_2O_3$ | Total | Log Kd[b] |
|---|---|---|---|---|---|---|---|---|---|---|---|---|---|---|
| E23 | r (5) | | FMQ-2 | Chr[c] | 14.71 (1.09)[d] | D.L. | 0.29 (4) | 12.04 (19) | 14.47 (16) | 0.39 (2) | 55.42 (1.03) | 0.32 (3) | 97.66 | 1.78 |
| E19 | r (5) | Chr-MORB | FMQ | Chr | 16.24 (1.68) | 0.09 (5) | 0.24 (4) | 16.82 (12) | 13.07 (33) | 0.41 (1) | 51.53 (2.17) | 0.15 (2) | 98.55 | 2.09 |
| E27 | r (14) | | FMQ+2 | Chr | 16.35 (3.21) | D.L. | 0.27 (3) | 16.49 (33) | 15.05 (91) | 0.36 (4) | 49.67 (4.25) | D.L. | 98.18 | 2.23 |
| E24 | r (5) | Chr-DiAn | FMQ-2 | Chr | 11.45 (1.65) | D.L. | D.L. | 1.73 (7) | 20.6 (21) | D.L. | 63.78 (1.84) | D.L. | 97.56 | 1.86 |
| E28 | r (10) | | FMQ+2 | Chr | 10.26 (1.57) | D.L. | D.L. | 1.06 (5) | 21.26 (23) | D.L. | 65.43 (2.08) | D.L. | 98.01 | 2.14 |
| E25 | c (2) | | FMQ-2 | Chr | 9.37 (1.97) | D.L. | 0.27 (5) | 13.27 (25) | 13.06 (32) | 0.36 (0) | 61.22 (1.98) | 0.19 (1) | 97.75 | 1.81 |
| E25 | r (5) | | FMQ-2 | Chr | 11.65 (0.49) | D.L. | 0.30 (03) | 13.25 (12) | 13.47 (23) | 0.41 (1) | 58.67 (0.51) | 0.27 (2) | 98.04 | 1.80 |
| E31 | c (12) | $Cr_2O_3$-MORB | FMQ | $Cr_2O_3$ | D.L. | D.L. | D.L. | 0.64 (28) | 0.18 (16) | 1.53 (3) | 97.48 (0.9) | 0.11 (1) | 99.94 | 2.34 |
| E31 | r (17) | | FMQ | Chr | 4.15 (3.26) | D.L. | 0.28 (4) | 16.52 (38) | 10.47 (1.14) | 0.45 (5) | 66.64 (4.59) | 0.1 (3) | 98.61 | 2.17 |
| E29 | c (5) | | FMQ+2 | $Cr_2O_3$ | 0.22 (3) | 0.14 (11) | D.L. | 1.08 (75) | 0.30 (27) | 1.55 (58) | 95.32 (2.69) | D.L. | 98.60 | 2.52 |
| E29 | r (15) | | FMQ+2 | Chr | 5.1 (3.07) | D.L. | 0.28 (5) | 18.56 (2.48) | 10.92 (1.37) | 0.34 (5) | 63.22 (7.14) | D.L. | 98.42 | 2.35 |
| E26 | c (3) | | FMQ-2 | MgChr | 4.77 (7) | D.L. | D.L. | 1.28(4) | 20.14 (7) | D.L. | 71.78 (7) | D.L. | 97.97 | 1.86 |
| E26 | r (5) | | FMQ-2 | MgChr | 7.72 (72) | D.L. | D.L. | 1.25(3) | 20.48 (2) | D.L. | 68.16 (92) | D.L. | 97.61 | 1.83 |
| E33 | c (19) | $Cr_2O_3$-DiAn | FMQ | $Cr_2O_3$ | 0.14 (1) | D.L. | D.L. | D.L. | 0.28 (28) | 1.45 (3) | 97.92 (1.19) | D.L. | 99.79 | 2.17 |
| E33 | r (12) | | FMQ | MgChr | 4.29 (2.68) | D.L. | D.L. | D.L. | 20.68 (46) | 0.14 (11) | 73.32 (3.93) | D.L. | 98.43 | 2.05 |
| E30 | c (14) | | FMQ+2 | $Cr_2O_3$ | 0.13 (13) | D.L. | D.L. | 0.21(6) | 0.31 (34) | 1.11 (53) | 98.13 (1.39) | 0.12 (1) | 100.00 | 2.28 |
| E30 | r (15) | | FMQ+2 | MgChr | 3.11 (2.05) | D.L. | D.L. | 2.72(33) | 19.31 (48) | 0.15 (11) | 73.35 (3.5) | D.L. | 98.64 | 2.16 |

[a] Analyses of central part of crystals (c) or rims (r) which are in equilibrium with melt. Number in brackets is a number of analyses.
[b] Logarithm of chromium distribution coefficient, ratio [concentration of $Cr_2O_3$ in crystal in wt% / concentration of $Cr_2O_3$ in melt in wt% ].
[c] Chr – chromite and MORB – starting MORB glass.
[d] Digits in brackets correspond to standard deviation of number of analyses.
[e] D.L. - value below detection limit.
'Phase' – means ____.

**Table 3. Composition of glasses in equilibrium with chromite or magnesiochromite (average values from multiple analyses)**

| Run[a] | System | Buffer | $Na_2O$ | $SiO_2$ | $Al_2O_3$ | $K_2O$ | CaO | MnO | FeO | MgO | $TiO_2$ | $P_2O_5$ | $Cr_2O_3$ | Total | Cr, ppm | Optical basicity |
|---|---|---|---|---|---|---|---|---|---|---|---|---|---|---|---|---|
| E23 (50) |  | FMQ-2 | 0.51 (3)[a] | 53.96(22) | 16.54 (1) | 0.19 (0.02) | 10.93 (1) | 0.18 (3) | 6.04 (9) | 8.98 (9) | 1.57 (3) | D.L.[b] | 0.92 (1) | 99.82 | 6306 (54) | 0.568 |
| E19 (50) | Chr-MORB | FMQ | 1.02 (4) | 52.13(17) | 15.88 (8) | 0.23 (0.02) | 10.67 (11) | 0.17 (4) | 8.91 (14) | 8.59 (8) | 1.50 (3) | D.L. | 0.42 (0) | 99.52 | 2856 (31) | 0.579 |
| E27 (50) |  | FMQ+2 | 1.84 (4) | 52.81(2) | 16.1 (11) | 0.27 (0.02) | 10.77 (11) | 0.17 (3) | 7.17 (12) | 8.64 (9) | 1.52 (3) | D.L. | 0.29 (0) | 99.58 | 1987 (26) | 0.578 |
| E24 (50) |  | FMQ-2 | D.L. | 49.77(2) | 9.21 (7) | D.L. | 24.66 (2) | D.L. | 1.04 (4) | 15.03 (11) | D.L. | 0.07 (2) | 0.88 (1) | 100.66 | 6023 (57) | 0.614 |
| E32 (50) | Chr-DiAn | FMQ | 0.11 (2) | 50.22(19) | 9.18 (7) | D.L. | 25.1 (14) | D.L. | 0.22 (3) | 15.11 (13) | D.L. | D.L. | 0.60 (1) | 100.54 | 4095 (45) | 0.615 |
| E28 (50) |  | FMQ+2 | 0.20 (2) | 50.2(19) | 9.12 (8) | D.L. | 25.01 (15) | D.L. | 0.51 (4) | 14.94 (11) | D.L. | 0.07 (2) | 0.48 (0) | 100.53 | 3260 (30) | 0.616 |
| E25 (50) |  | FMQ-2 | 0.60 (3) | 54.66(17) | 15.90 (10) | 0.20 (0.02) | 11.08 (1) | 0.16 (3) | 6.27 (11) | 8.31 (8) | 1.63 (3) | D.L. | 0.94 (1) | 99.75 | 6415 (53) | 0.567 |
| E31 (50) | $Cr_2O_3$-MORB | FMQ | 1.26 (4) | 52.90(13) | 15.41 (1) | 0.25 (0.02) | 10.78 (11) | 0.17 (3) | 8.54 (11) | 8.00 (8) | 1.58 (3) | D.L. | 0.45 (1) | 99.34 | 3058 (52) | 0.576 |
| E29 (50) |  | FMQ+2 | 2.16 (4) | 51.66(18) | 15.04 (9) | 0.27 (0.02) | 10.59 (1) | 0.16 (3) | 9.74 (11) | 7.80 (7) | 1.54 (3) | D.L. | 0.28 (0) | 99.24 | 1948 (26) | 0.583 |
| E26 (49) |  | FMQ-2 | D.L. | 50.61(17) | 8.46 (8) | D.L. | 25.17 (17) | D.L. | 0.68 (4) | 14.47 (12) | 0.06 (2) | D.L. | 1.00 (1) | 100.45 | 6824 (56) | 0.612 |
| E33 (50) | $Cr_2O_3$-DiAn | FMQ | 0.10 (2) | 51.15(21) | 8.55 (6) | D.L. | 25.55 (18) | D.L. | D.L. | 14.55 (13) | D.L. | D.L. | 0.66 (0) | 100.56 | 4498 (32) | 0.613 |
| E30 (50) |  | FMQ+2 | 0.13 (2) | 50.30(2) | 8.38 (7) | D.L. | 25.15 (16) | D.L. | 1.59 (6) | 14.24 (8) | 0.06 (2) | D.L. | 0.51 (1) | 100.36 | 3500 (35) | 0.616 |

Notes:

[a] Number in brackets is a number of analyses.

[a] Digits in brackets correspond to standard deviation from average value for the performed analyses.

[b] D.L. - value below detection limit.

# Appendix 1. The script for fitting of the new thermodynamic model written on Python language.

```
Last login: Mon Jul 24 07:40:09 on console
MacBook-Pro-Polzovatel:~ polzovatel$ Python
Python 2.7.10 (default, Oct 23 2015, 19:19:21)
[GCC 4.2.1 Compatible Apple LLVM 7.0.0 (clang-700.0.59.5)] on darwin
Type "help", "copyright", "credits" or "license" for more information.
>>>
>>> import os
>>> os.getcwd()
'/Users/polzovatel'
>>> os.chdir('/Users/polzovatel/Desktop/Python')
>>> os.getcwd()
'/Users/polzovatel/Desktop/Python'
>>> import numpy as np
>>> from scipy.optimize import curve_fit
>>> x,y,z,m = np.loadtxt('Real data2.txt', skiprows=0, unpack=True)
>>> def func(X, a, b, c, d, e, f):
...     x,y,z=X
...     return np.exp(a+b*z+c/x)*(1+(1/(y**(0.25)))*np.exp(d+e/x+f*z))
...
>>> print curve_fit(func, (x,y,z), m)
(array([ -7.01016484e+00,   1.37185711e+01,  -1.24049202e+04,
         2.44577629e+01,  -2.43946573e+04,  -2.35912896e+01]), array([[  4.42743908e+00, -5.43107435e+00,  -1.94807684e+03,
         -6.12726816e+00,   2.91048455e+03,   7.30579890e+00],
       [ -5.43107435e+00,   7.45148747e+00,   1.64100727e+03,
          7.64424126e+00,  -2.39730802e+03,  -1.04343461e+01],
       [ -1.94807684e+03,   1.64100727e+03,   1.57107470e+06,
          2.60163410e+03,  -2.39291743e+06,  -1.87073714e+03],
       [ -6.12726816e+00,   7.64424126e+00,   2.60163410e+03,
          9.79566379e+00,  -4.32049314e+03,  -1.21132562e+01],
       [  2.91048455e+03,  -2.39730802e+03,  -2.39291743e+06,
         -4.32049314e+03,   4.15074890e+06,   2.91585605e+03],
       [  7.30579890e+00,  -1.04343461e+01,  -1.87073714e+03,
         -1.21132562e+01,   2.91585605e+03,   1.76145858e+01]]))

>>>
```